\documentclass[traditabstract]{aa}
\usepackage{graphicx}
\usepackage{txfonts}
\usepackage{natbib}
\usepackage{longtable}
\usepackage[footnotesize]{caption}

\def\figpath{Figures/}

\begin{document}
\title{Neptune Trojans and Plutinos: colors, sizes, dynamics, and their possible collisions}

\titlerunning{Neptune Trojans and Plutinos: colors, sizes, dynamics, and collisions}

\author{A.J.C. Almeida\inst{1,2} \and N. Peixinho\inst{3,4} \and A.C.M.
Correia\inst{1,5}}

\authorrunning{A.J.C. Almeida et al. }


\institute{Departmento de F\'{\i}sica, Universidade de Aveiro, Campus de Santiago,
3810-193 Aveiro, Portugal
   \and
           Instituto de Telecomunica\c{c}\~oes, IT - Aveiro, Campus de Santiago,
3810-193 Aveiro, Portugal
   \and
           Center for Computational Physics, University of Coimbra, Portugal
   \and
           Astronomical Observatory of the University of Coimbra, Portugal
   \and
           Astronomie et Syst\`emes Dynamiques, IMCCE-CNRS UMR8028, 
77 Av. Denfert-Rochereau, 75014 Paris, France
           }

\date{Received 24 February / Accepted 23 September 2009}

\abstract{
Neptune Trojans and Plutinos are two sub-populations of Trans-Neptunian Objects
located in the 1:1 and the 3:2 mean motion resonances with Neptune, respectively, and 
therefore protected form close encounters with the planet.
However, the orbits of these two kinds of objects may cross very often, allowing
a higher collisional rate between them than with other kinds of Trans-Neptunian Objects 
and a consequent and size distribution alteration of the two sub-populations.

Observational colors and absolute magnitudes of Neptune Trojans and Plutinos 
show that: i) there are no intrinsically bright (large) Plutinos at small inclinations; 
ii) there is an apparent excess of blue and intrinsically faint (small) Plutinos; 
and iii) Neptune Trojans possess the same blue colors as Plutinos within the same
(estimated) size range do.

For the present sub-populations we analyze the most favorable conditions for close
encounters / collisions to occur and address if there 
could be a link between those encounters and the sizes and/or colors of 
Plutinos and Neptune Trojans.
We also perform a simultaneous numerical simulation of the outer Solar System
over 1~Gyr for all these bodies in order to estimate their collisional rate.

We conclude that orbital overlap between Neptune Trojans and Plutinos is
favored for Plutinos with high libration amplitudes, high eccentricities and
low inclinations. 
Additionally, with the assumption that the collisions can be disruptive 
creating smaller objects not necessarily with similar colors, 
the present high concentration of small Plutinos at low inclinations can thus be a
consequence of a collisional interaction with Neptune Trojans and such hypothesis should be 
further analyzed.
}


\maketitle


\section{Introduction}

Trans-Neptunian Objects (TNOs), also known as Kuiper Belt Objects (KBOs), are a population of small and
primitive icy bodies orbiting (mostly) beyond Neptune. Their study is one of the most significant ways to
obtain information on the early ages of the Solar System.  

Based on some distinct dynamical properties, the TNOs can be subdivided in several different sub-populations, 
often also called families or groups.
Subdivide them as a function of  their physical properties seems to be far more complex. TNOs have surface
colors so diverse that can go from blue/neutral ({\it i.e.} solar-like) to extremely red.  
A possible explanation for the wide variety of colors was originally proposed by
\cite{1996AJ....112.2310L} with the collisional resurfacing model. In this model, the competition between
surface reddening, due to cosmic-ray bombardment, and a resurfacing with frozen material (assumed to be
bluer) withdrawn from beneath its crust by impact collisions could be responsible for the observed wide
range of surface colors. As the model seems to fail, at least in its simple form, more complex forms have
been proposed. Namely, collisional resurfacing combined with cometary activity \citep{Delsanti+04}, and
collisional resurfacing with layered reddening and even re-bluing by cosmic-rays
\citep{2002P&SS...50...57G}. An alternate idea proposes that surface colors are primordial \citep[][and
references therein]{TRC03}. 
More recently, \cite{2009Icar..199..560G} shows that an object could lose its redness simply by ice sublimation 
(without resurfacing). Such could explain the lack of red colors among TNOs that evolved into Jupiter family 
comets due to their closeness to the Sun. Yet, as the aforementioned work acknowledges, the existence of 
the blue/neutral TNOs at heliocentric distances where ices do not sublimate suggests the existence of other 
coloring mechanisms. Our understanding on the origin and eventual alteration of TNOs colors is
still very limited and, up to the present, none of these approaches lead to a fully
consistent explanation for the color diversity \cite[for a review see][]{2008ssbn.book...91D}. 

Among the TNOs, two sub-populations caught our attention: the Neptune Trojans and the Plutinos. 
Neptune Trojans are the small bodies trapped in a 1:1 mean motion orbital resonance with Neptune, also
with orbital eccentricities lower than 0.1. Roughly they co-orbit with Neptune concentrated $60^{\circ}$
ahead and $60^{\circ}$ behind Neptune's position. 
\cite{2005ApJ...628..520C} proposed that Neptune Trojans formed by in-situ accretion from small-sized debris 
once Neptune's migration has stopped. 
On the other hand, other works indicate they can survive planetary migration 
\citep{2002Icar..160..271N, 2004Icar..167..347K} and, more recently, due to the discovery of a large population 
of high inclination Neptune Trojans, \cite{2009AJ....137.5003N} sustain they were captured during planetary migration
Compared to TNOs Neptune Trojans seem to be quite small in
size (diameters $D<100$ km) and also slightly blue. 

At the time of our analysis 6 Neptune Trojans were known\footnote{See
http://cfa-www.harvard.edu/iau/lists/NeptuneTrojans.html}, all of them librating around the Lagrangian
point $L_4$. 

Plutinos are those trapped in a 3:2 mean motion orbital resonance
with Neptune, with eccentricity values ranging between 0.1 and 0.3. Unlike Trojans, the
colors of Plutinos vary from blue/neutral to the very red, and their sizes range from a few tens
of~km to a few thousands (note that Pluto is a Plutino). Roughly, Plutinos possess semi-major axes within
$39 < a < 40.5$ AU. 
Through long-term dynamical evolution studies \cite{2007Icar..189..213L} identified 98 Plutinos and that will
be our reference.

Being locked at the 3:2 resonance, Plutinos can periodically cross the orbit of Neptune without colliding
with it. However, this protection from collisions is not possible for Neptune Trojans and Plutinos. 
A first look at the geometry of these two populations suggests that they might even collide frequently. 

The analysis of the collisional resurfacing model by \cite{2003Icar..162...27T} and \cite{2003EM&P...92..233T} found that Plutinos were
significantly more affected by collisions than other TNOs. Since, observationally, Plutinos did not
exhibit bluer colors than all other TNOs taken as a whole, the aforementioned work strongly argued against the
collisional resurfacing model, at least as major cause for the color diversity of TNOs. 
Neptune Trojans were not included in \citeauthor{2003Icar..162...27T}'s simulations, though,  
since they were not known at the time. 
Notwithstanding the arguments against it, the collisional resurfacing that was analyzed was a 
very simple model. For instance, the possibility of collisional disruption of differentiated bodies 
producing fragments or ruble-piles with surface colors distinct from those of the parent bodies has not been 
taken into account in the collisional resurfacing models yet. Naturally, the possible outcomes of this scenario 
would be far more complex and our understanding of the physics of collisional processes between the icy TNOs is 
still limited \citep[see review by][]{2008ssbn.book..195L}. Nonetheless, both a collisional family of TNOs associated 
with (136108) Haumea and a collisional system of satellites associated with Pluto have been detected 
\citep{2007Natur.446..294B, 2006Natur.439..946S, 2005Sci...307..546C}.

Surveys show a power-law for the size distribution of TNOs with slope change at
$D\sim100$~km  \citep{2004AJ....128.1364B}, which is most consistent with TNOs being gravity dominated
bodies with negligible material strength. Objects larger than $\sim 100$~km are difficult to
disrupt, hence likely primordial, and objects smaller than this are expected to be shattered due to
disruptive collisional evolution \citep{2008ssbn.book..293K, 2005Icar..173..342P}. 

\cite{Elia+08} studied
the collisional evolution of Plutinos considering only Plutino-Plutino collisions.  
They infer that any eventual slope change at  $D\sim40-80$~km on the power-law 
size distribution of Plutinos should be primordial and not of collisional origin. Yet, they find that
collisional populations may form from the breakup of objects larger than $100$~km, and if those populations form
at low inclinations their fragments will likely stay in the resonance. 

On Sect. \ref{obs} we will analyze the observational
data relative to Neptune Trojans and Plutinos. We will discuss 
two particular observational properties: 
1) there is an apparent ``excess'' of small blue Plutinos and these happen to be also in the same size 
range and to possess the same colors as the known Neptune Trojans, and 2) there is
a concentration of small Plutinos, and absence of large ones, at low orbital inclinations. 
The eventual collisional interaction between Plutinos and Neptune Trojans has not been analyzed in 
previous works. Since their geometry points to the existence of mutual 
collisions, the existence of 
an eventual link between the colors and sizes of Neptune Trojans and Plutinos and their
interactions/collisions motivated us to this work. On Sect. \ref{dynamiceq} we will
study the dynamics of Neptune Trojans and Plutinos, separately, focusing on their orbits around the
Lagrangian point $L_4$. On Sect. \ref{numsim} we will discuss the possibility of 
collisions between Neptune Trojans and Plutinos and the best conditions for this to happen.
Finally, on Sect. \ref{conclusions} we will present our conclusions.

\section{Observational Results and Discussion}
\label{obs}

\begin{figure}[h!]
\centering
\includegraphics[width=0.45\textwidth]{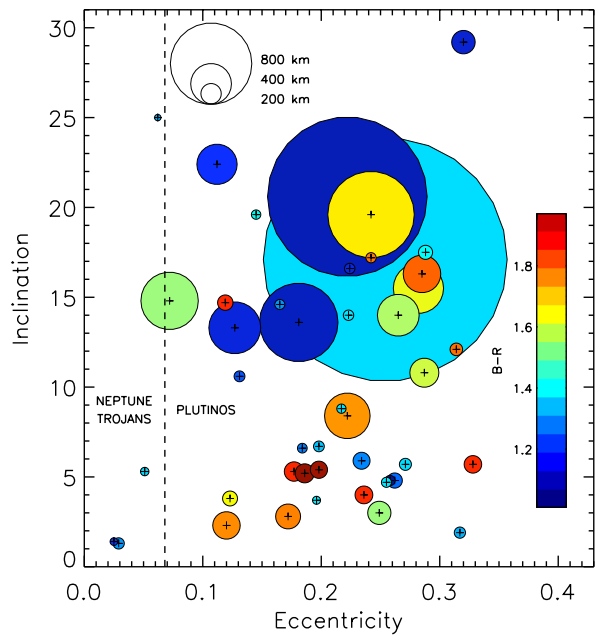}
\caption{Orbital inclination vs eccentricity, estimated size, and color of Neptune Trojans and Plutinos
for which those properties have been measured.} 
\label{i_e_plot_plut_ntroj_new}
\end{figure}

In Table~\ref{tab:tab_1} we summarize the orbital elements, $B-R$ colors, and R-filter
absolute magnitudes ($H_R$) for 4 Neptune Trojans \citep[from][]{2006Sci...313..511S} and 41 Plutinos (see
references on the table). 
In Fig. \ref{i_e_plot_plut_ntroj_new} we plot the orbital inclination vs orbital eccentricity of all these
objects, together with their $B-R$ colors, indexed on a color palette on the right hand side of the figure, and
in which objects are plotted proportionally to their estimated diameter also indexed on the top of the figure.

The diameters of these objects (in kilometers) are estimated from the absolute magnitudes using the formula
\citep{1916ApJ....43..173R}: 
\begin{equation}
D_{(km)} = 2\times \sqrt{\frac{2.24\times 10^{16} \times 10^{0.4(-27.10 - H_R)}}{p_R}}\,,
\end{equation}
where $H_R$ is the R-filter absolute magnitude, and the R-filter albedo is taken as $p_R=0.09$
\citep{2004AJ....127.2413B}. A size estimation exception is
made for Pluto which is represented with $D=2390$~km. A vertical dashed line separates 
Neptune Trojans, located on the left side of the figure,
from Plutinos, located on the right side.  

A first look at Fig. \ref{i_e_plot_plut_ntroj_new} shows that:
\begin{description}
\item[(i)] all Trojans are blue\footnote{For simplicity throughout this work we will call an object blue
when $B-R<1.5$ and red when $B-R\geq1.5$. The $B-R$ color of the Sun is 1.03.}, small ($D<100$ km)
compared to the size distribution of Plutinos, and possess small eccentricities;
\item[(ii)] there is an apparent concentration of small Plutinos at low inclination values and a
concentration of large Plutinos at high inclinations; 
\item[(iii)] Plutinos have larger eccentricity values than Trojans, and their colors range from blue to
red appearing randomly distributed in inclination and eccentricity.  
\item[(iv)] all Plutinos within the same (estimated) size range as Neptune Trojans possess blue colors.
\end{description}

It is important to note that the small sized Plutinos, which are also as blue as Neptune Trojans, are randomly
scattered in eccentricity and inclination whereas the low inclined Plutinos are all relatively small but
range from blue to red colors. 

Based on these properties we decide to explore two possible scenarios:
 
\begin{description}
\item[(1)] could the equally blue colors of the equally sized Plutinos and Neptune Trojans be 
the result of some collisional interaction between both populations? 
\item[(2)] could the concentration of small Plutinos at low inclinations be the result of some 
collisional interaction between them and Neptune Trojans? 
\end{description}

Why these two scenarios? From the simulations by \cite{2003Icar..162...27T} and \cite{2003EM&P...92..233T}, which analyzed the 
collision rates among all TNOs simultaneously (except for Neptune Trojans that were not known at the time), 
Plutinos received more collisions than any other sub-population. The number/energy of those 
collisions also seemed independent of their orbital parameters. That is,  Plutinos seemed under the same collisional 
environment regardless of their high or low inclination and/or eccentricity values. 

Therefore, since the possible effects of a significant number of Neptune Trojans were not considered in the 
aforementioned simulations in this work we explore if and how these Trojans could be the major cause of the 
non-homogeneity of surface properties observed among Plutinos and the (apparent) homogeneity of 
surface properties observed for Neptune Trojans.

Let us grasp scenario (1). The members of the collisional families associated with (136108) Haumea
possess very similar bluish colors \citep{2007Natur.446..294B}.
It is reasonable to hypothesize that if Neptune Trojans
collided heavily with part of the Plutino population the collision outcomes could be 
similar in color. Considering that assumption, 
for this scenario to be possible we need to find a similar collision rate between Trojans and Plutinos 
independent of the eccentricity and/or inclination values of Plutinos since the small ($D<100$
km) and blue Plutinos are homogeneously scattered both in eccentricity and inclination.  

As to scenario (2), we are considering a less strict hypothesis in which the collision outcomes of  
collisions between Neptune Trojans 
and part of the Plutino population are not necessarily equal in color.
The Pluto system has, presumably, a collisional origin, however, while Charon, Nix, and Hydra possess 
similar colors Pluto is quite distinct \citep[e.g.][]{2009Icar..199..571S}. 
Let us assume that the disruption of large layered TNOs could 
generate several objects with distinct colors either from fragments coming from different layers with different colors 
or from reaccretion of different parts of a heterogenous cloud of debris. Hence, 
collisions would generate both blue and red, small and medium objects ($D<300$ km).
For this scenario to be possible we need to find a much higher
collision rate between Trojans and the Plutinos with low inclination values than with the Plutinos at high
inclinations. 
Also, given that we do not see neither red nor mid/large-sized Neptune Trojans, whereas we see it among 
Plutinos in the presumable region of collisions, the primordial size 
and color distributions of Trojans and Plutinos should not have been similar.

Note that even though resonant objects are known to possess large variations of their eccentricity and/or inclination values,
\cite{NesvornyRoig00} showed the existence of two dynamical populations of Plutinos, separating at $i\sim10^{\circ}$, which 
do not seem to easily mix. Consequently, any different collisional/surface evolution that both populations might have had in the 
past should still be detectable today.

\begin{figure}[h!]
\centering
\includegraphics[width=0.5\textwidth]{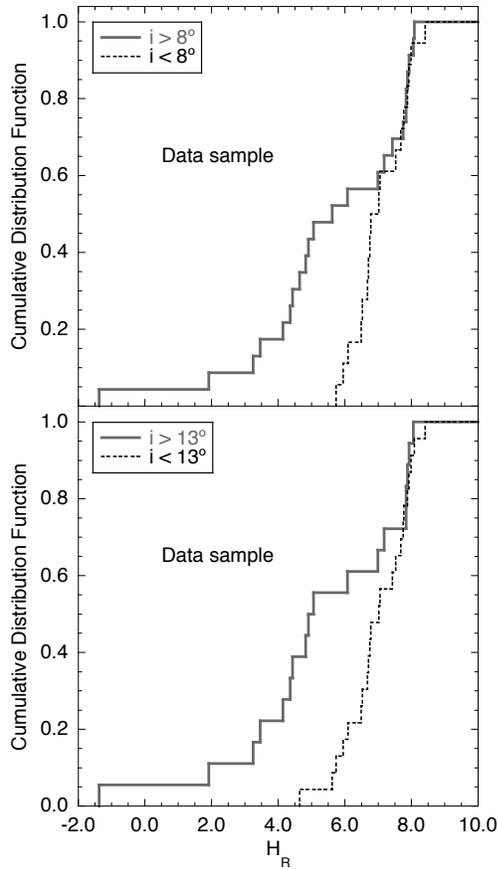}
\caption{Cumulative distribution function of absolute magnitudes $H_R$ for Plutinos above and below
$8^{\circ}$ of inclination (top) and Plutinos above and below $13^{\circ}$ (bottom). The incompatibility
of distributions between the higher and lower inclined Plutinos is similar in both cases. There is an
evident  lack of intrinsically bright (large) objects among the lower inclined Plutinos.} 
\label{cumul_distrib_pluts}
\end{figure}

In order to test if the low inclined Plutinos are indeed smaller than the other ones we use the
Kolmogorov-Smirnov test \citep{Press+92, Kolmogorov33, Smirnov39}.
We successively compare the absolute magnitude $H_R$ distribution of our 41 Plutinos (with measured $B-R$
colors) having $i<i_c$ with those 
having $i\geq i_c$, varying $i_c$ from $4^{\circ}$ to $15^{\circ}$ in increments of $0.5^{\circ}$. Two
significant solutions exist: a) Plutinos with $i< 8^{\circ}$ are intrinsically fainter, and assumed to be
smaller, than those above $8^{\circ}$ with a (one-tailed) significance $p=0.0024$; b) Plutinos with $i<
13^{\circ}$ are intrinsically fainter than those above $13^{\circ}$ with a significance $p=0.0030$. Even
though we do not obtain an unique solution for the best inclination value that separates the larger
(brighter) from the smaller (fainter) Plutinos, we can conclude that the low-inclined ones are in fact
smaller. Figure \ref{cumul_distrib_pluts} shows the cumulative distribution functions of absolute magnitudes
for the higher inclined and lower inclined Plutinos in both solutions. 

We will proceed with the study of the dynamics of Plutinos and Neptune Trojans and investigate their
possible collisions.

\section{Dynamical Evolution}
\label{dynamiceq}

In previous section we discussed the possible relations between the colors of
Plutinos and Neptune Trojans, and the eventual collisions between them.
It is now our goal to test the possibility of collisions numerically.
For that purpose we will simulate the outer Solar System evolution, where the
TNOs are considered massless. This hypothesis is essential to speed up the
integrations.
The equation of motion for planets is given by
\begin{equation}
\ddot{\textbf{r}}_i + \mathcal{G}(m_s+m_i)\frac{\textbf{r}_i}{r_i^3}= \sum_{j \neq i}^{N_P} \mathcal{G}m_j \Bigg(\frac {\textbf{r}_j-\textbf{r}_i}
{\left|\textbf{r}_j - \textbf{r}_i\right|^3} - \frac {\textbf{r}_j} {r_j^3} \Bigg)\,,
\label{Eq1}
\end{equation}
where $ \textbf{r}_i $ is the vector position of the planet, $\mathcal{G}$ the
gravitational constant, $m_s$ the mass of the Sun, $m_i$ the mass of the
planet, and $N_P$ is the total number of planets.
In our simulations we will only take into account the four giant planets.
The effect of the inner Solar System in the dynamics of the Kuiper belt objects
is only residual and by neglecting it we may use a larger step-size for
numerical simulations and considerably decrease the length of the simulations.

For TNOs, since they are assumed massless, the equation of motion is given by 
\begin{equation}
\ddot{\textbf{r}}_k + \mathcal{G}m_s \frac{\textbf{r}_k}{r_k^3}= \sum_{j}^{N_P} \mathcal{G}m_j \Bigg(\frac {\textbf{r}_j-\textbf{r}_k}
{\left|\textbf{r}_j - \textbf{r}_k\right|^3} - \frac {\textbf{r}_j} {r_j^3} \Bigg) \,,
\label{Eq2}
\end{equation}
where $ \textbf{r}_k $ is the vector position of the TNO.
By adopting the above equations we assumed that planets and TNOs are only
perturbed by the remaining planets, i.e., the TNOs are considered as test
particles.
The only exception will be the Pluto-Charon system barycenter, which will be
considered as a planet because of its important mass. Indeed, 
some Plutinos may be pushed out of the 3:2 resonance
by Pluto into close encounters with Neptune 
\citep{1999AJ....118.1873Y, Nesvorny_etal_2000}.

In our simulations we will use the symplectic integrator from
\cite{2001CeMDA..80...39L}, with an integration step-size of 0.1\,yr.
The system is composed of 5 planets (Table\,\ref{tab:tab_2}), 6 Trojans
(Table\,\ref{tab:tab_3}) and 98 Plutinos (Table\,\ref{tab:tab_4}).

\subsection{Resonant Motion}

Since both Trojans and Plutinos are resonant objects we will briefly recall here
the bases of resonant motion, following \cite{Murray_Dermott_1999}.
Consider then a TNO in some resonance with Neptune. We assume for simplicity,
that Neptune is in a circular orbit and that all motion takes 
place in the plane of its orbit. We also ignore any
perturbations between the two bodies as we are only interested in how 
resonant relationships lead to repeated encounters.

We can examine the geometry of resonance for a general case by first considering
two bodies moving around the Sun, in circular and 
coplanar orbits. So, let us assume that
\begin{equation}
\label{meanmotion}
\frac{n'}{n} = \frac{p}{p+q}\,,
\end{equation}
where \emph{n} and \emph{n'} are the mean motions of Neptune and the TNO,
respectively (for Trojans $p = 1$ and $q = 0$, while for Plutinos $p = 2$
and $q = 1$). If the two bodies are in conjunction at time $t=0$, the next
conjunction will occur when 
$(n - n')t = 2\pi$, and the period, $T_{con}$, between successive conjunctions is given by
\begin{equation}
T_{con} = \frac{2\pi}{n - n'} = \frac{p}{q}T' = \frac{p+q}{q}T\,,
\end{equation}
where \emph{T} and \emph{T'} are the orbital periods of the two bodies.

Now consider the case when $e=0$, $e' \neq 0$, and $\dot \varpi' \neq 0$, where \emph{e} denotes the eccentricity for Neptune and
\emph{e'} and $\varpi'$ denotes the eccentricity and longitude of pericentre of
the TNO respectively. If the resonant relation
\begin{equation}
\label{eq:relres}
(p+q)n' - pn - q\dot \varpi' = 0
\end{equation}
is satisfied, then we can rewrite Eq.(\ref{meanmotion}) as
\begin{equation}
\frac{n' - \dot \varpi'}{n - \dot \varpi'} = \frac{p}{p+q}\,,
\end{equation}
where $n' - \dot \varpi'$ and $n - \dot \varpi'$ are relative motions. These can
be considered as the mean motions in a reference frame, 
co-rotating with the pericentre of the TNO. From the point of view of this
reference frame, the orbit of the TNO is fixed or stationary. 
If the resonant relation given in Eq.(\ref{eq:relres}) holds, the corresponding
resonant argument is 
\begin{equation}
\label{eq:angress}
\varphi = (p+q) \lambda' - p\lambda - q\varpi' \,,
\end{equation}
where $\lambda$ and $\lambda'$ denote the mean longitude of Neptune and the
TNO, respectively. At a conjunction of the two bodies, 
$\lambda = \lambda'$ and we have
\begin{equation}
\varphi = q(\lambda' - \varpi') = q(\lambda - \varpi')\,.
\end{equation}
Thus, $\varphi$ is a measure of the displacement of the longitude of conjunction
from pericentre of the TNO.
By computing the derivative of the resonant angle $\varphi$, we get
\begin{equation}
\label{eq:dangress}
\dot \varphi = (p+q) n' - pn - q\dot \varpi'\,,
\end{equation}
and $ \dot \varphi = 0 $ from  Eq.(\ref{eq:relres}).
In a more general situation, we will have $ \dot \varphi \neq 0 $, but in order
to preserve the resonant equilibrium $ \varphi $ will librate around an
equilibrium position $ \varphi_0 $, obtained when $ \dot \varphi = 0 $.
The libration amplitude $ \Delta \varphi $ will depend on the initial conditions
and perturbations from the other bodies in the system and may reach large
values.
As a consequence, it is possible that the orbits of two distinct bodies librating
around different equilibrium positions intercept at some point.

Along with the orbital parameters of the Trojans and Plutinos that we used in
our simulations, in Table\,\ref{tab:tab_3} and Table\,\ref{tab:tab_4}, respectively, we provide the equilibrium
libration angle, the main libration period and amplitude of each TNO
obtained over the next 250~Kyr.

\subsection{Trojans}
\label{trojansapp}

\begin{figure}[h]
\begin{center}$
\begin{array}{cc}
\includegraphics[scale=0.25, angle=0]{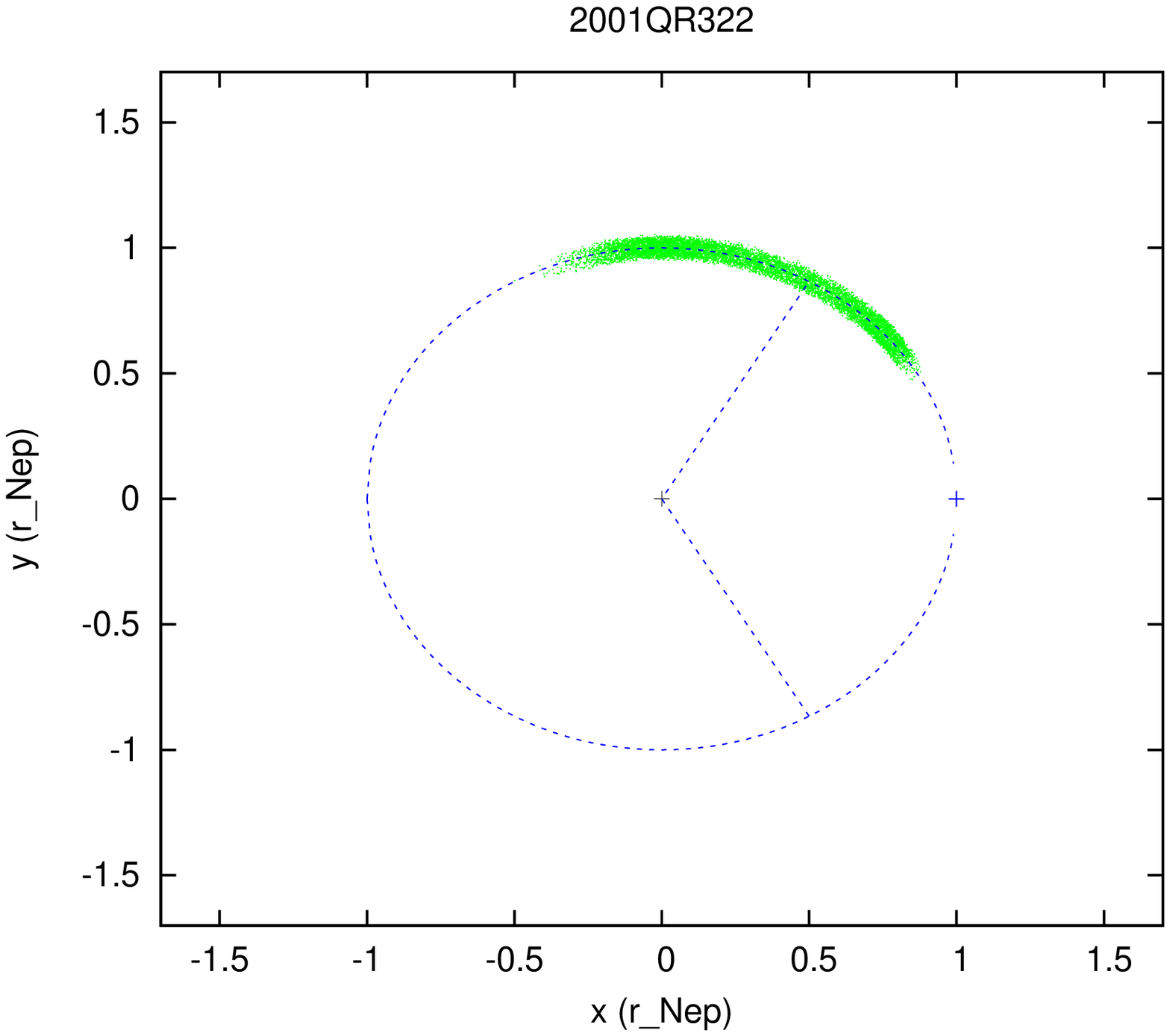} &
\includegraphics[scale=0.25, angle=0]{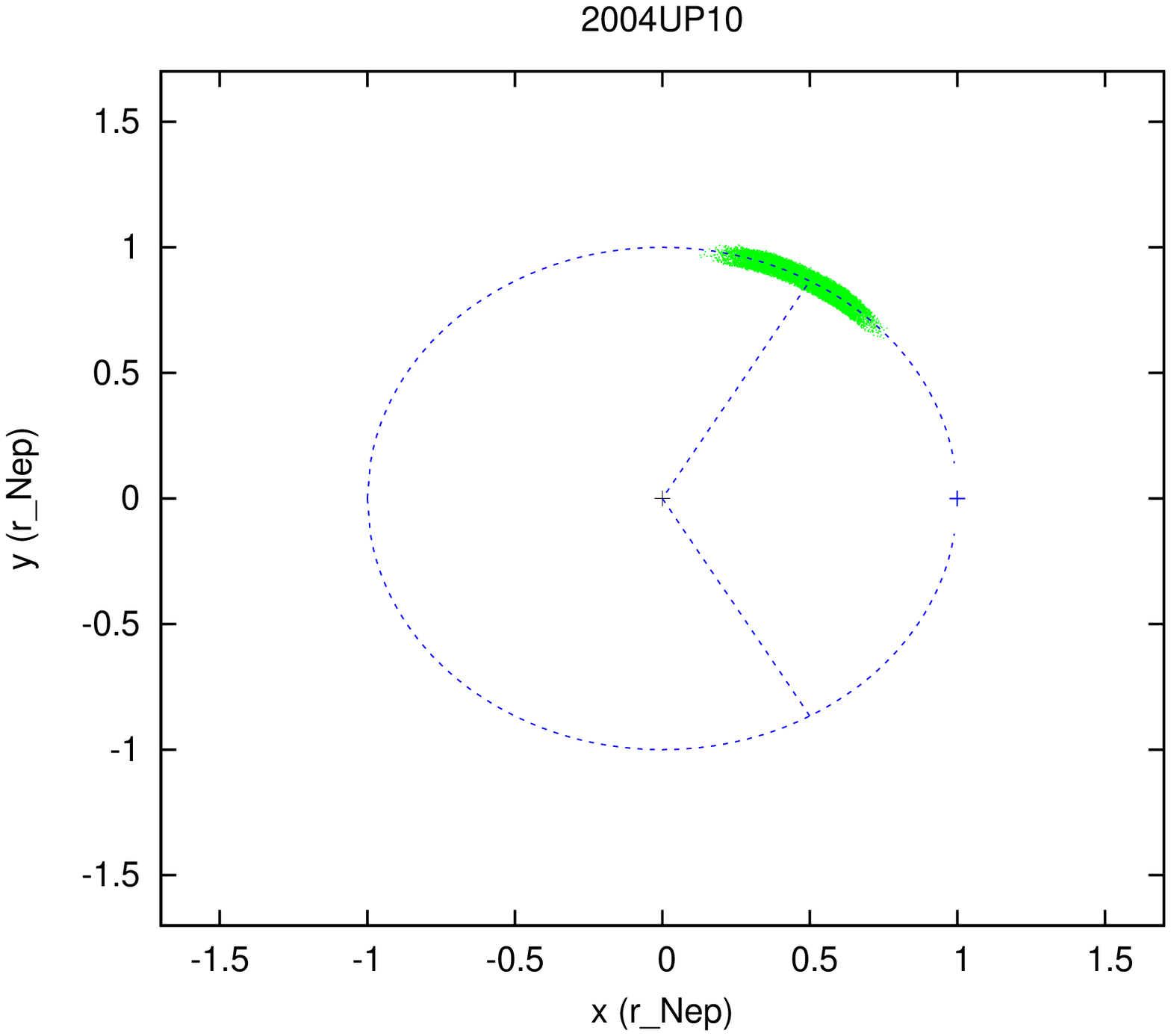} \\
\includegraphics[scale=0.25, angle=0]{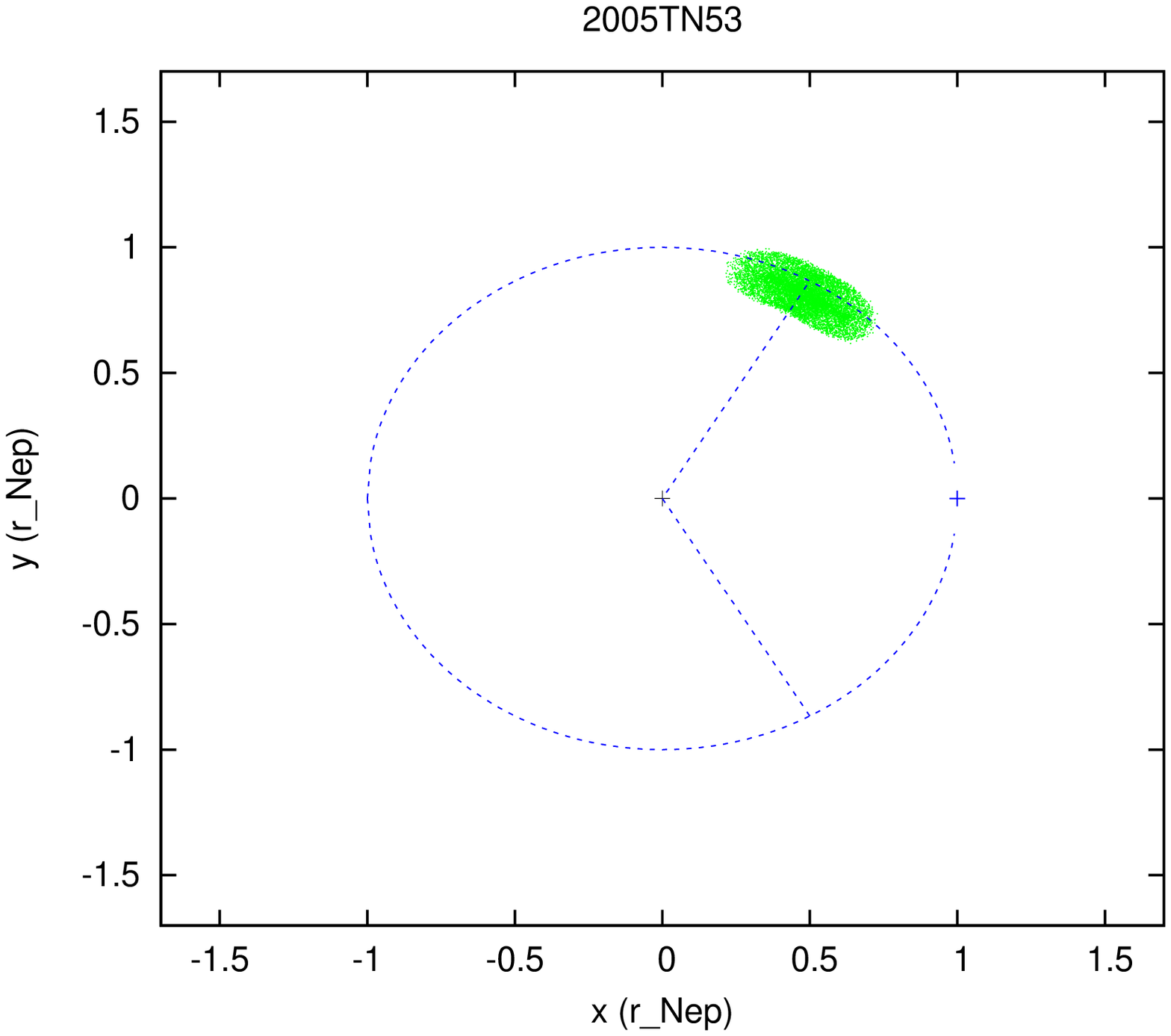} &
\includegraphics[scale=0.25, angle=0]{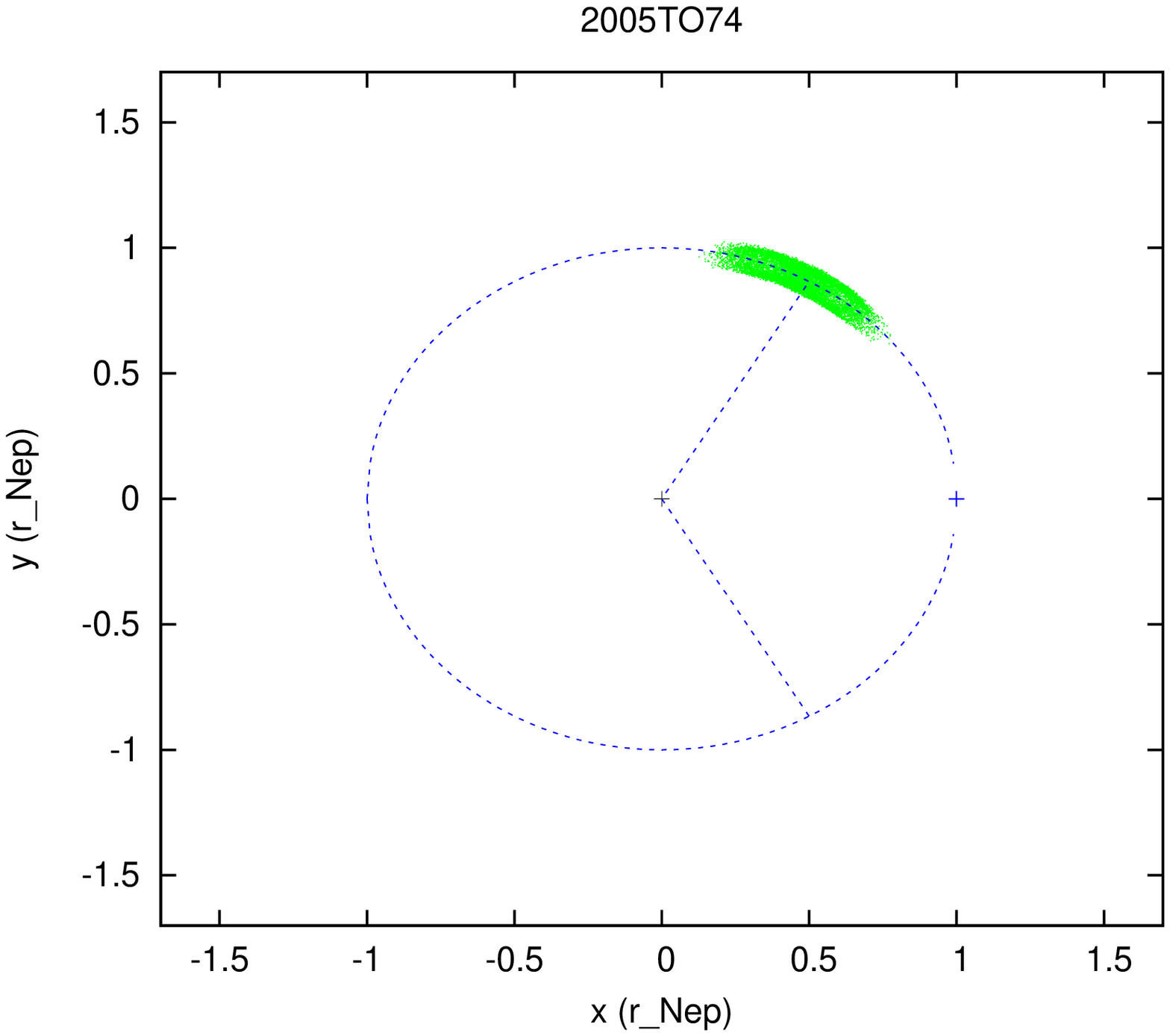} \\
\includegraphics[scale=0.25, angle=0]{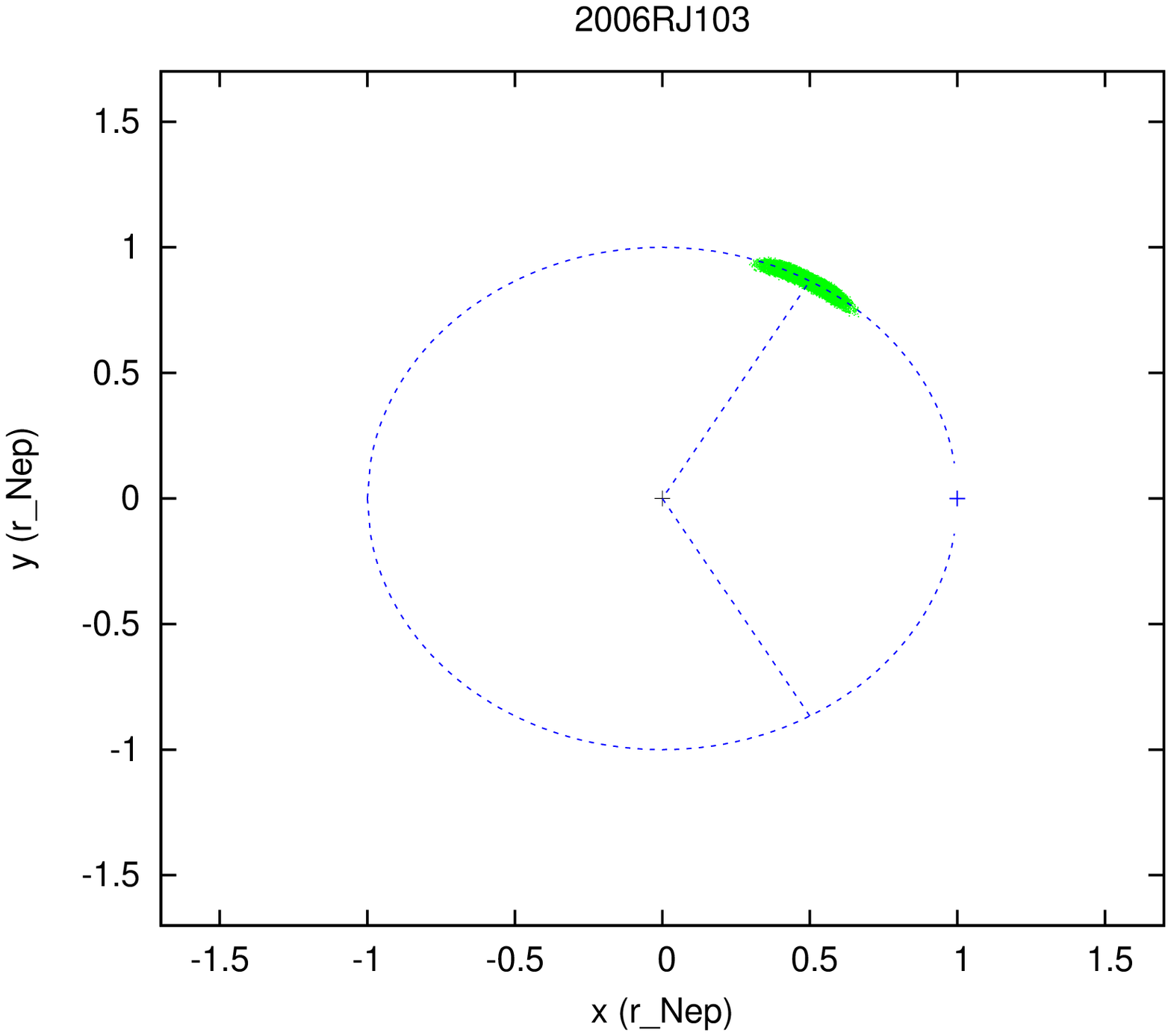} &
\includegraphics[scale=0.25, angle=0]{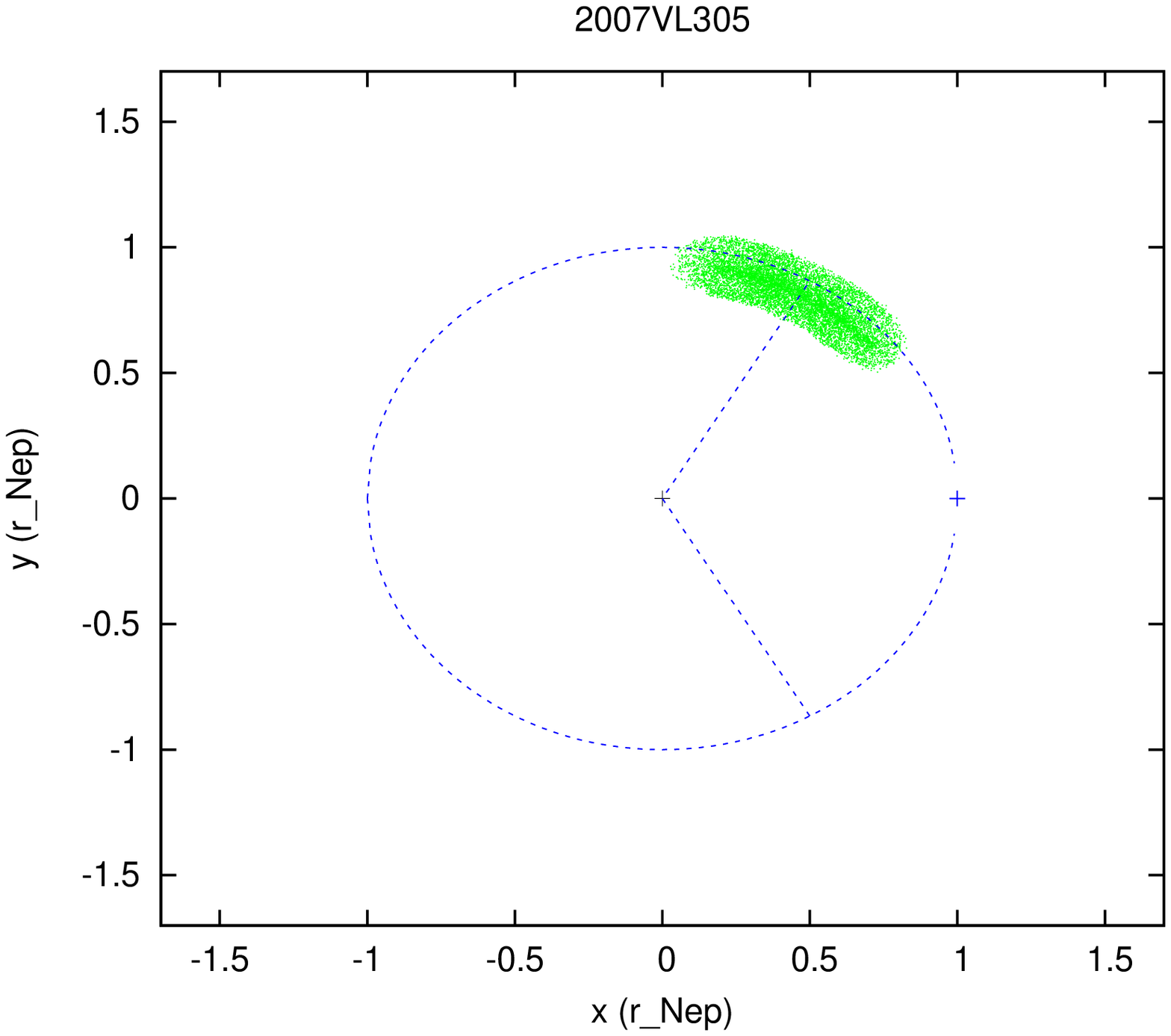}
\end{array}$
\end{center}
\caption{Orbital evolution of the Neptune Trojans (in green) listed in Table\,\ref{tab:tab_3}
over 100~Myr in the co-rotating frame of Neptune (in blue). Each panel shows the
projection of the TNO position every 10~Kyr in the
orbital plane of Neptune. $x$ and $y$ are
spatial coordinates centered in the star and rotating with Neptune, normalized
by the Neptune-Sun distance. All Trojans orbit around the Lagrangian point
$L_4$, and execute tadpole-type orbits. More scattered orbits correspond to higher
values of the eccentricity and inclination, while distance to the $L_4$ point
depend on the libration amplitude.}  
\label{trojans}
\end{figure}

Neptune Trojans are resonant objects in a 1:1 mean motion resonance
with Neptune. In our model we computed the motion of the six Neptune
Trojans listed in Table\,\ref{tab:tab_3}.
In Fig. (\ref{trojans}), we show the behavior of all known Trojans, along time, in a
co-rotating frame with Neptune for 100~Myr. 
Each dot shows the position of the TNO every 10\,kyr.

As expected, we see in Fig.\ref{trojans} that all Trojans orbit around the
Lagrangian point $L_4$, and execute tadpole-type orbits. 
This kind of orbits represents stable oscillations 
in the vicinity of the Lagrangian equilibrium point 
\citep[e.g.][]{2007PhyD..225..112G}.
The differences between the shape of their orbits, depend
on the libration amplitude, but also on their orbital eccentricity 
and inclination values. Trojan 2007VL305, that execute the most
scattered orbit, also presents the largest eccentricity and inclination
($e = 0.062$ and $i = 28^\circ$).
On the other hand, for small values of these two orbital parameters, the
TNOs remain roughly in the path of Neptune's orbit, only changing its
relative position to the planet due to libration.

In Table\,\ref{tab:tab_3} we provide the libration amplitude and period for all Trojans.
While amplitudes can vary from only $6^\circ$ to $26^\circ$, the periods of
libration remain around 9\,Kyr for all objects.
Trojan 2001QR322 presents the largest libration amplitude and therefore moves
further away of the equilibrium point $L_4$.
As a consequence, its orbit will be more susceptible of being destabilized by
gravitational perturbations from the planets and other bodies in the system.
Indeed, in our long-term numerical simulations (Sect.~\ref{stability}) this TNO will
abandon the Trojan orbit after 112\,Myr and become a Kuiper belt object.

\subsection{Plutinos}
\label{Plutinos}

\begin{figure}[h]
\centering
\includegraphics[width=.45\textwidth]{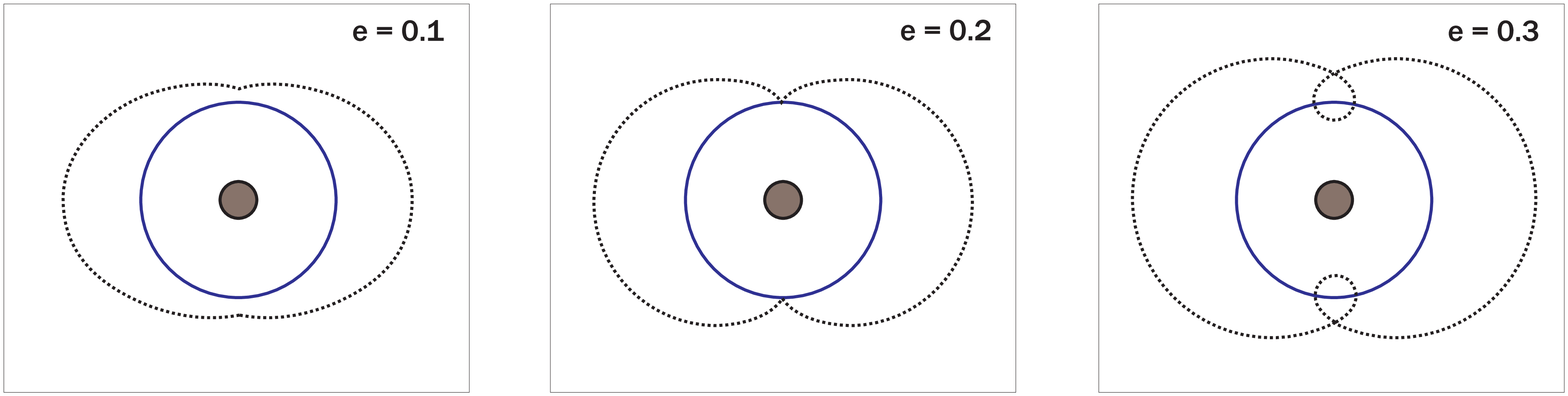} 
\caption{Typical path of a Plutino (dotted line) in the rotating frame of
Neptune (full line) for different eccentricity values ($ e = 0.1$, 0.2 and 0.3).
The position of the Plutino was drawn for equal time intervals.
Only high eccentricity values ($e > 0.2 $) allow the Plutino to cross the orbit
of Neptune. Due to the 3:2 mean motion resonance the trajectories are repeated
every two orbits of the TNO around the Sun.} 
\label{plutinopaths}
\end{figure}

Plutinos are resonant TNOs in a 3:2 mean motion resonance with Neptune. 
Thus, like Trojans, although they can cross the orbit of Neptune, they are
protected from possible encounters with this planet. 
In Fig.\ref{plutinopaths} we have drawn the typical path of a Plutino in the
co-rotating frame of Neptune, for three different values of the eccentricity ($ e =
0.1$, 0.2 and 0.3). 

\begin{figure}[h]
\centering
\includegraphics[width=.3\textwidth]{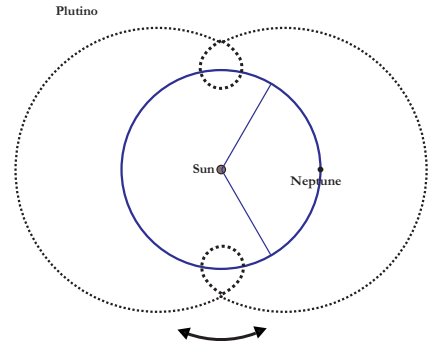}
\caption{Libration motion of the orbit of a Plutino.}
\label{lib_plut}
\end{figure}

These plots are drawn assuming that the Plutino is at exact resonance ($ \dot
\varphi = 0 $), which is not true, because the orbit is librating
around an equilibrium position $\varphi_0$ (Eq.\ref{eq:angress}).
As a consequence, in a more realistic situation we will observe an oscillation of
those paths as the one represented in Fig.\ref{lib_plut}.
In Table\,\ref{tab:tab_4} we provide the libration amplitude and period for all Plutinos.
The equilibrium libration angle for them all is $ \varphi_0 = \pm 180^\circ
$, but the amplitudes of libration can be as small as $ 7^\circ $ for Plutino
1996TQ66 or as wide as $ 120^\circ $ for Plutinos 1995QY9 and 2001KN77.
The libration periods, $ P_{lib} $,  vary between 14.5 and 26.6~Kyr, the average 
being around 20~Kyr.
For comparison, the values for Pluto are $ \Delta \varphi = 79.7^\circ$ and $
P_{lib} = 19.9$~Kyr. 

In our model we computed the motion of about 100 Plutinos, whose orbital
parameters are listed in Table\,\ref{tab:tab_4}.
All objects present moderate eccentricities and inclinations, ($\bar{e} \sim
0.23$ and $\bar{i} \sim 10.4^\circ$).
According to \citet{1995AJ....110..420M} these values can be a consequence of the
resonant mechanism of capture, during the residual planetesimal cleaning in the
vicinity of the young giant planets. Due to Neptune's migration, the
eccentricity and inclination of the Plutinos are pumped after capture in
resonance.

\begin{figure}[h]
\begin{center}$
\begin{array}{cc}
\includegraphics[scale=0.25, angle=0]{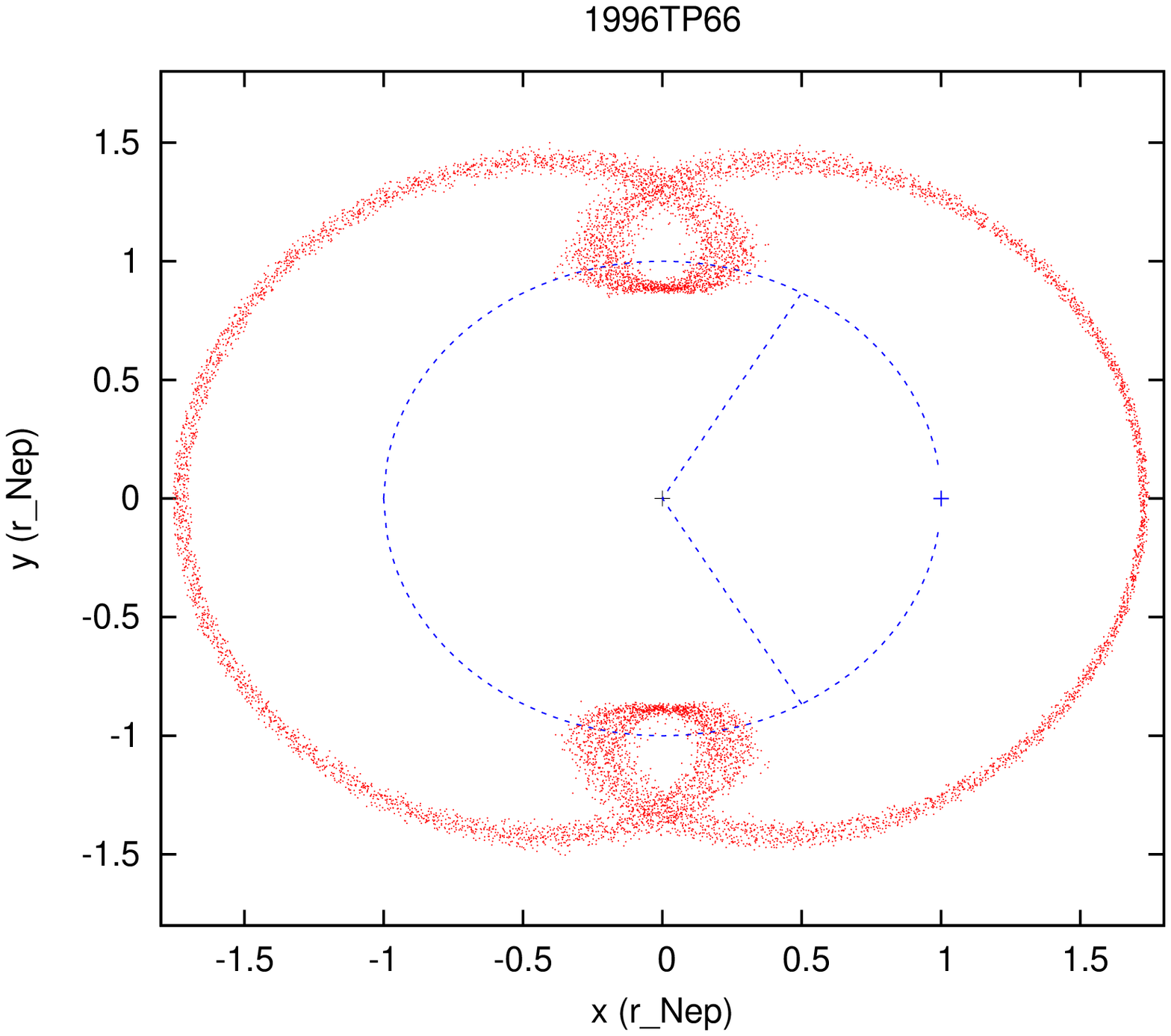} &
\includegraphics[scale=0.25, angle=0]{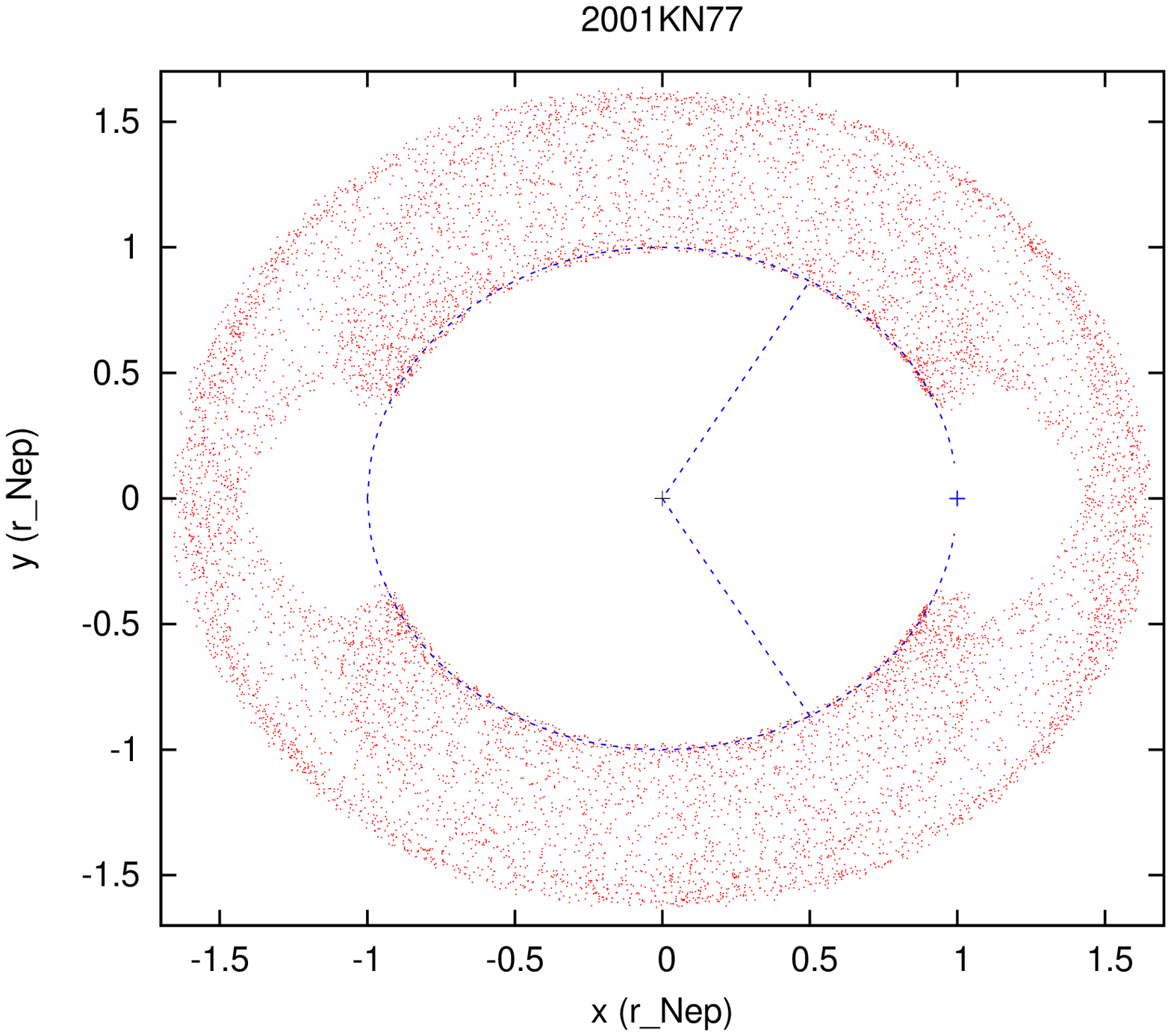} \\
\includegraphics[scale=0.25, angle=0]{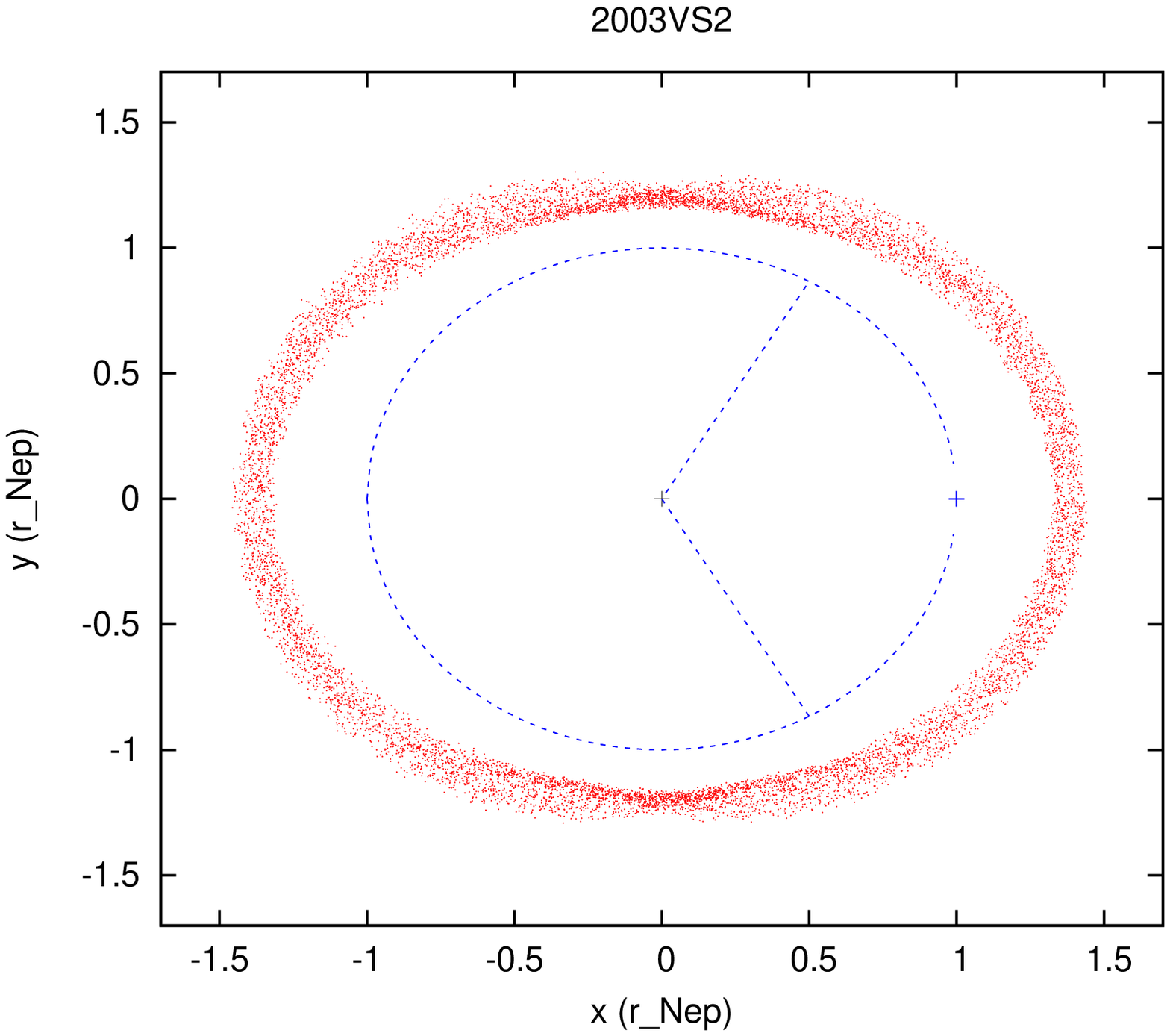} &
\includegraphics[scale=0.25, angle=0]{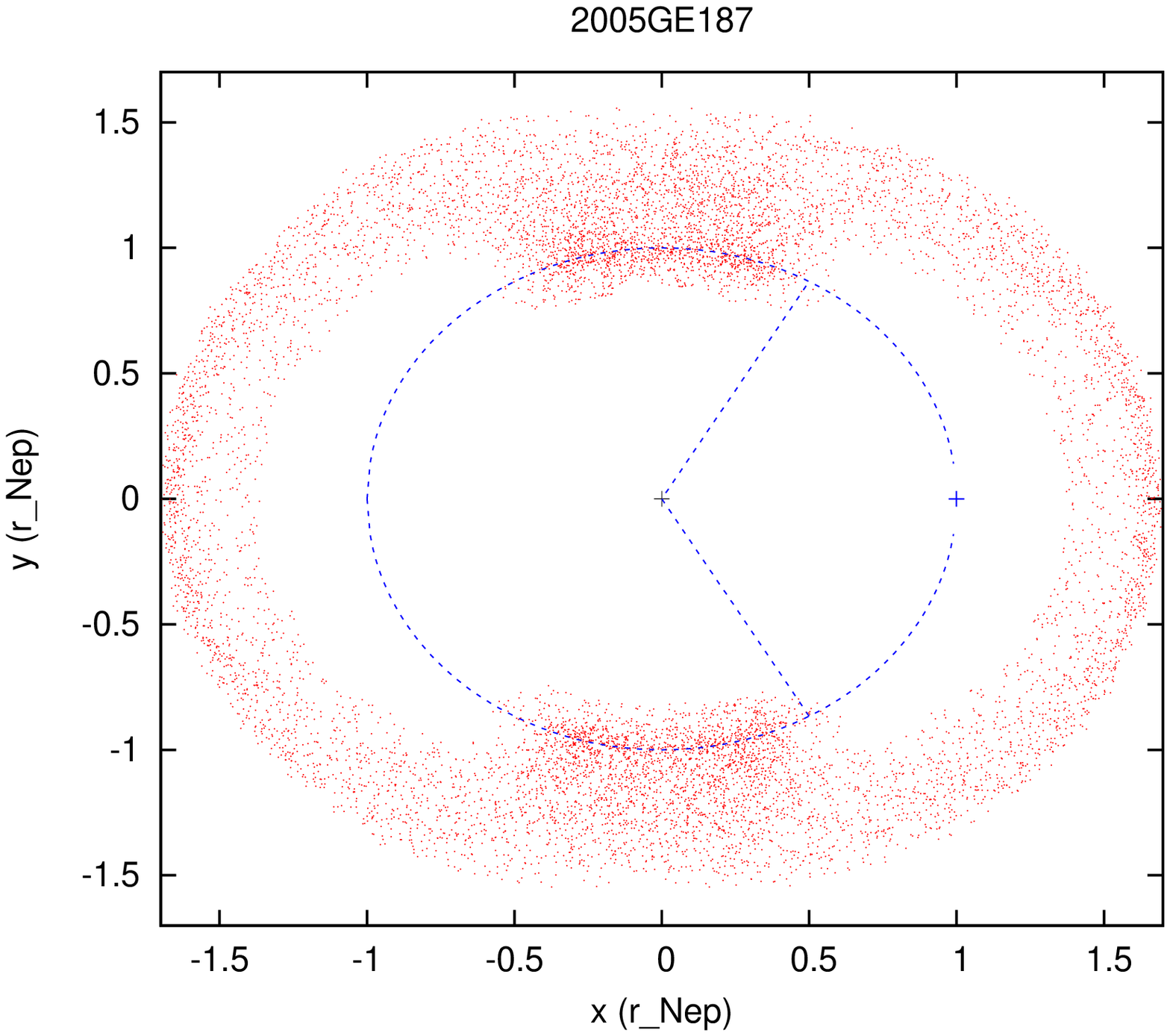} \\
\includegraphics[scale=0.25, angle=0]{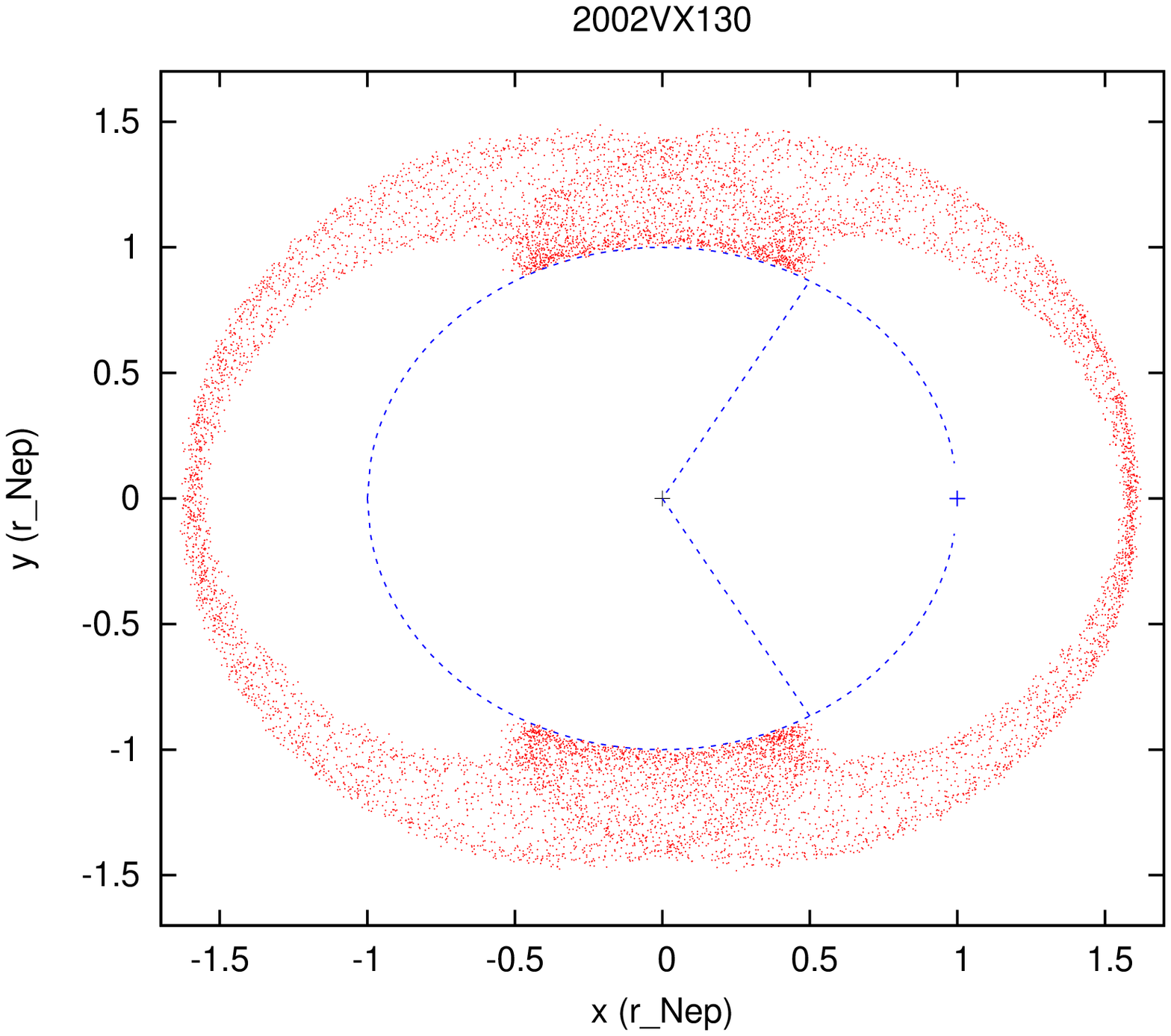} &
\includegraphics[scale=0.25, angle=0]{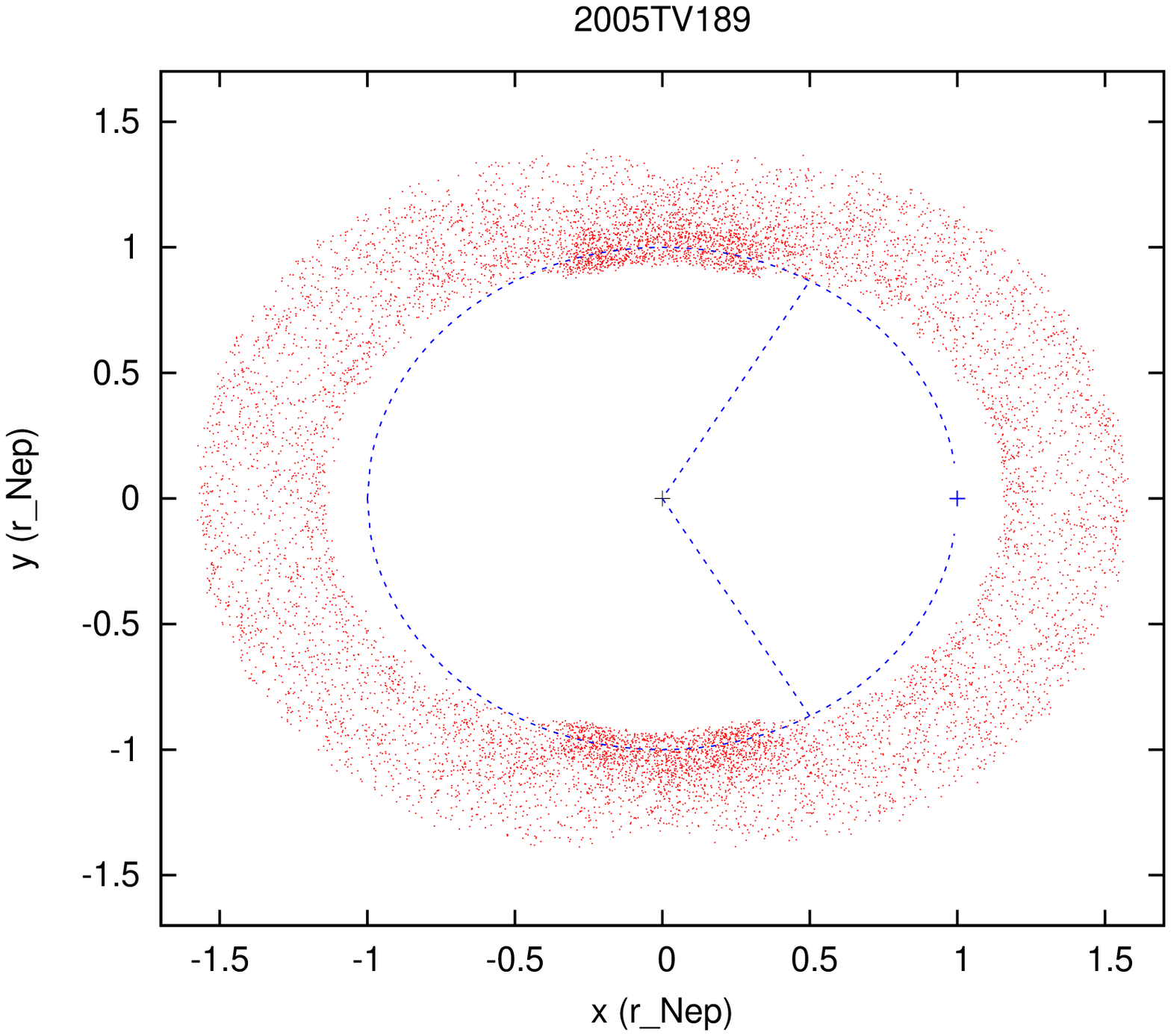}
\end{array}$
\end{center}
\caption{Orbital evolution of some Plutinos (in red) taken from
Table\,\ref{tab:tab_4} over 100~Myr in the co-rotating frame of Neptune (in
blue). Each panel shows the projection of the TNO position every 10~Kyr in
the orbital plane of Neptune. $x$ and $y$ are
spatial coordinates centered on the star and rotating with Neptune, normalized
by the Neptune-Sun distance. 
We have chosen the most representative cases, namely, Plutino 1996TP66, with
the smallest libration amplitude ($\Delta \varphi = 7.2^\circ$, $e=0.328$,
$i=5.6^\circ$), Plutino 2001KN77, with the widest 
libration amplitude ($\Delta \varphi = 120.4^\circ$, $e=0.242$, $i=2.4^\circ$), 
Plutino 2003VS2, with the smallest eccentricity ($e=0.072$, $i=14.8^\circ$),
 Plutino 2005GE187, with the highest eccentricity ($e=0.329$, $i=18.2^\circ$), 
 Plutino 2002VX130, with the smallest inclination ($e=0.220$,
$i=1.3^\circ$), and Plutino 2005TV189, 
with the highest inclination ($e=0.186$, $i=34.5^\circ$).
}
\label{plutinos}
\end{figure}

In Fig.\ref{plutinos} we show the behavior of six Plutinos,
along time, in a co-rotating frame with Neptune for 100~Myr, each dot showing
the position of the object every 10\,Kyr. 
Because Plutinos are much more numerous than Trojans, we can only represent a
small fraction of them. We have chosen the most representative cases,
namely, the Plutino with the smallest and widest libration (1996TP66 and
2001KN77, respectively), the Plutinos with smallest and highest eccentricity
(2003VS2 and 2005GE187, respectively), and the Plutinos with smallest and 
highest inclination (2002VX130 and 2005TV189, respectively).

As expected, depending on the eccentricity values, the orbits of the Plutinos
are all in good agreement with the paths shown in Fig.\ref{plutinopaths}.
Because of the libration motion of the orbits (Fig.\ref{lib_plut}) their
trajectories approach or cross the Lagrangian points $L_4$ and $L_5$, that is,
the Plutinos orbits share the same spatial zone as the Neptune Trojans.

\section{Numerical Simulations}
\label{numsim}

In Sect.\ref{obs} we have shown that, although Neptune Trojans and Plutinos appear to have
different origins, some of their properties indicate that the two types are
quite identical, suggesting that there must be some sort of
communication between them.
Indeed, in the previous section we saw that due to the libration motion of the orbits there
is a wide zone of spatial overlap around the
Lagrangian point $L_4$ of Neptune.
As a consequence, we may expect close encounters and collisions between the two kinds 
of TNOs to occur at a higher rate than in the remaining Kuiper belt, 
resulting in a change of the size distributions of the two populations.

In order to test this possibility we simulated the long-term future evolution of
the outer Solar System for 1~Gyr.
The orbits of the outer planets, the Neptune Trojans and the Plutinos are
integrated simultaneously according to the model described in the beginning of
Sect.\ref{dynamiceq}.

\subsection{Stability of the Neptune Trojans}
\label{stability}

The stability of the Neptune Trojans orbits is an important issue on the
dynamics of the outer Solar System. 
According to \citet{2007MNRAS.382.1324D} Trojans with low-inclined orbits are
less stable.
The stability area around $L_4$ and $L_5$ disappears after about $10^8$~yr for
low inclinations, while this stability zone is still present for about
$10^9$~yr for large inclinations.
More precisely, it was concluded that there exists a region ($20^\circ < i <
50^\circ$) of higher stability for the Neptune Trojans, although only two
have presently been found in this region (Table\,\ref{tab:tab_3}).

During our numerical simulations all Trojans remained stable within the limits 
shown in Fig.\ref{trojans} during 1~Gyr, except 
Trojan 2001QR322, which escapes
from the Lagrangian $L_4$ point after about 112~Myr.
This last observation was somehow unexpected, since \citet{2003AJ....126..430C}
concluded that this same Neptune Trojan was stable over 1~Gyr.
This difference of behaviors is a consequence of a modification in the initial
conditions (Table\,\ref{tab:tab_3}), obtained with new observational data.
This event is more or less in agreement with the results from
\citet{2007MNRAS.382.1324D}, because Trojan 2001QR322 is the one with smallest
inclination  ($ i = 1.3^\circ $). 
Although other Trojans with identical inclination values remained stable,
we notice that Trojan 2001QR322 also has the widest libration amplitude ($
\Delta \phi = 26^\circ $) and its orbit is therefore more susceptible of being
disturbed by planetary perturbations.
This suggests that the stability of the Trojans is smaller for low-inclined
orbits, but also for large libration amplitudes.

\begin{figure}[h]
\centering
\includegraphics[scale=0.35, angle=-90]{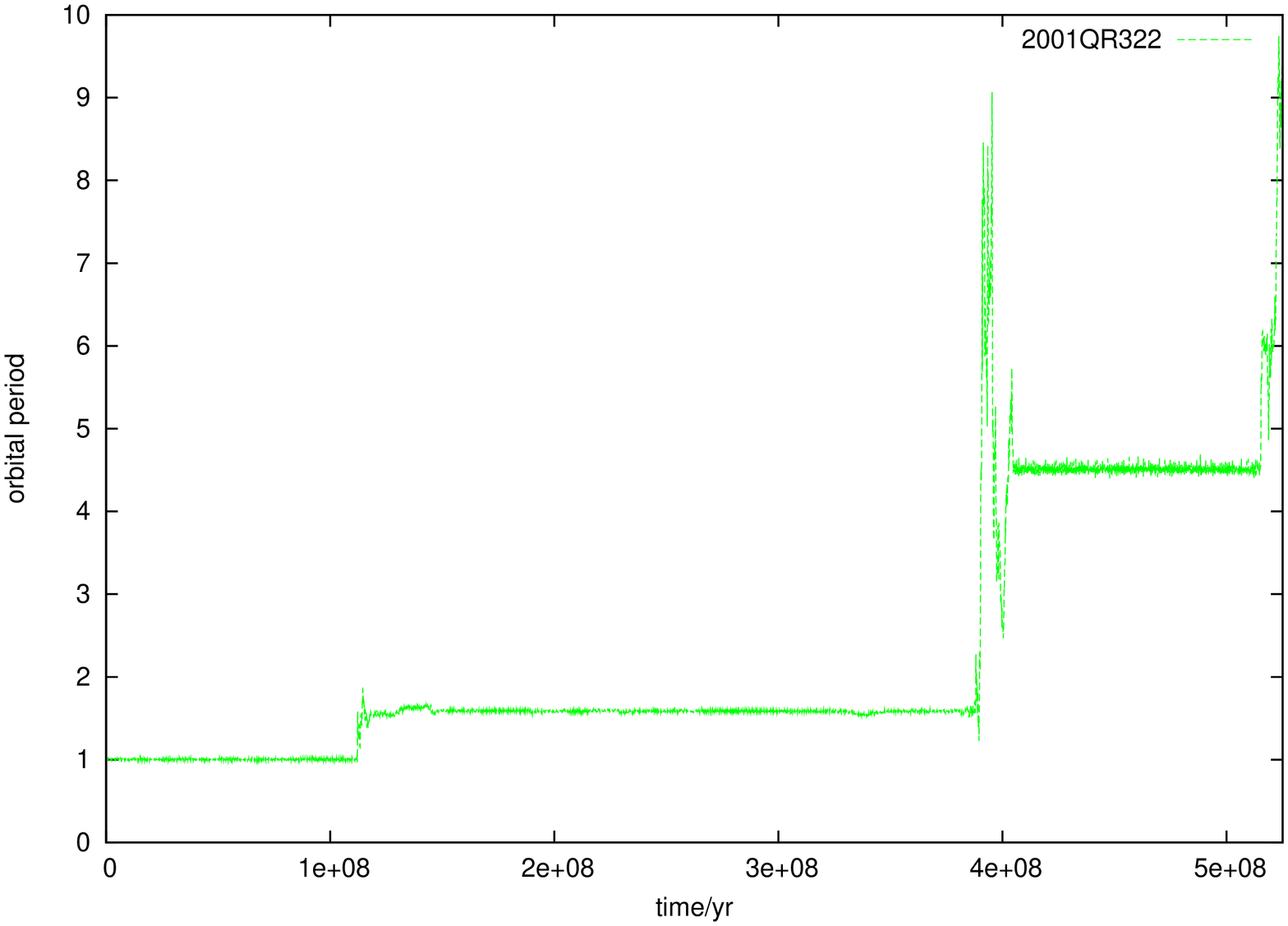}\\
\includegraphics[scale=0.35, angle=-90]{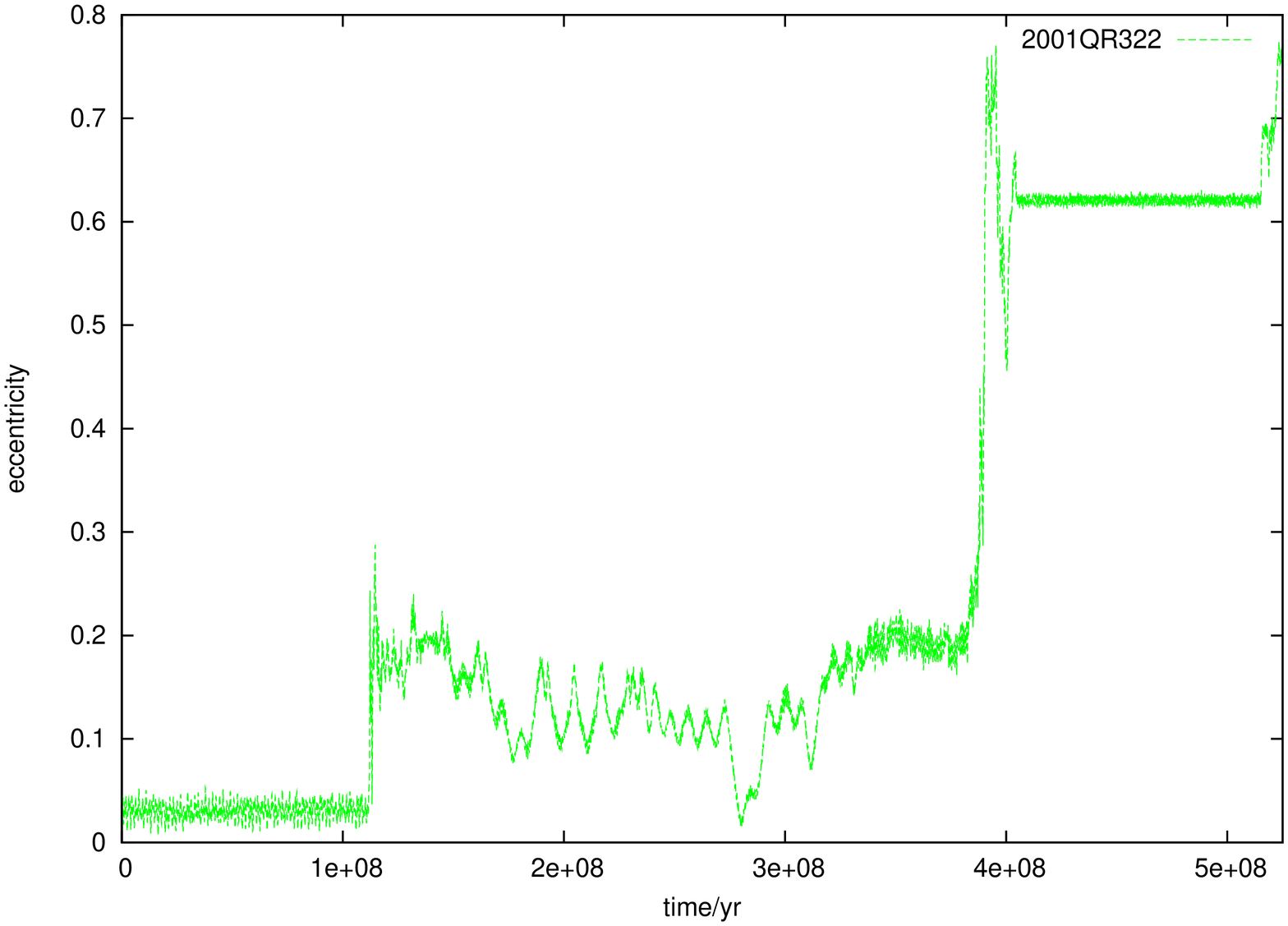}
\caption{Long-term evolution of the orbital period (over Neptune's orbital
period) and the eccentricity of Trojan 2001QR322 for 525~Myr.
Initially a Trojan, the orbit of this TNO is not stable and quits the 1:1
mean motion resonance after about 112~Myr. Then it jumps to different
configurations until it becomes a comet and eventually collides with a planet or
the Sun. In this simulation it is ejected from the Solar System after 525~Myr.} 
\label{2001qr322_orb_ecc}
\end{figure}

In Fig.\ref{2001qr322_orb_ecc} we plot the long-term evolution of the orbital
period and the eccentricity of the Trojan 2001QR322.
The behavior of this object is extremely regular, until it undergoes a sudden
increment of the eccentricity, which removes it form the 1:1 mean motion
resonance with Neptune.
Interestingly, just after escaping the resonance, the semi-major axis of
this TNO is temporarily stabilized near the 3:2 resonance with Neptune,
i.e., its orbit becomes very close to the Plutino's orbits.
However, contrary to regular Plutinos, the eccentricity undergoes important
chaotic variations from nearly zero up to 0.3.
This regime lasts for about 300~Myr, time after which the eccentricity grows to
almost 0.8.
As a consequence, close encounters with the planets become possible and the
TNO leaves the area near to the 3:2 mean motion resonance.
Later on, the same TNO seems to be captured in a 9/2 mean motion resonance
with Neptune, and stays there for about 100~Myr.
Finally, the orbit is again destabilized and the eccentricity grows to extreme
values, and the TNO turns into a comet.

\subsection{Stability of the Plutinos}
\label{stabilityP}

The orbits of the Plutinos are believed to be stable as long as
the libration amplitudes are smaller than $120^\circ$ \citep{NesvornyRoig00}.
For larger libration widths, the planetary perturbations will
remove the TNO from its orbit in short time intervals.
The same phenomena was observed for Trojans.

During our numerical simulations over 1~Gyr, 10 Plutinos over 98
quit their orbits (about 10\%). 
In Table\,\ref{tab:tabZ} we list all these bodies, as well as their libration
width (Table\,\ref{tab:tab_4}).
Contrary to expectations, we observe that only two Plutinos present libration
amplitudes around $120^\circ$.
The minimum libration width observed is around $40^\circ$.
A possible explanation is that because Plutinos are not alone in their orbits and, 
according to \citet{1999AJ....118.1873Y} and \citet{Nesvorny_etal_2000}, some Plutinos may be pushed out of the
resonance by Pluto into close encounters with Neptune.
Indeed, stability maps by \citet{NesvornyRoig00} only take into account the effect of
the four giant planets.

\begin{table}[t!]
  \caption{List of all unstable Plutinos during 1~Gyr.
  \label{tab:tabZ}}   
  \begin{tabular}{l c c r}
\hline
\hline 
\textbf{Plutino} & \textbf{time (Myr)} & \textbf{$e$} & \textbf{$\Delta \varphi \, (^\circ)$} \\ 
\hline
 2000FV53  & 69 &0.168 & 117.76  \\
 2004FU148 &230 &0.235 &  94.67  \\
 2002GE32  &500 &0.232 &  73.14  \\
 1998WZ31  &590 &0.165 &  76.61  \\
 2004EW95  &595 &0.320 &  53.31  \\
 1995QY9   &680 &0.262 & 120.22  \\
 2003TH58  &710 &0.088 &  65.29  \\
 1993SB    &720 &0.317 &  53.19  \\
 2002XV93  &820 &0.127 &  42.70  \\
 2003UT292 &840 &0.292 &  80.69  \\
\hline
\end{tabular}
\end{table}

\begin{figure}[h]
\centering
\includegraphics[scale=0.35, angle=-90]{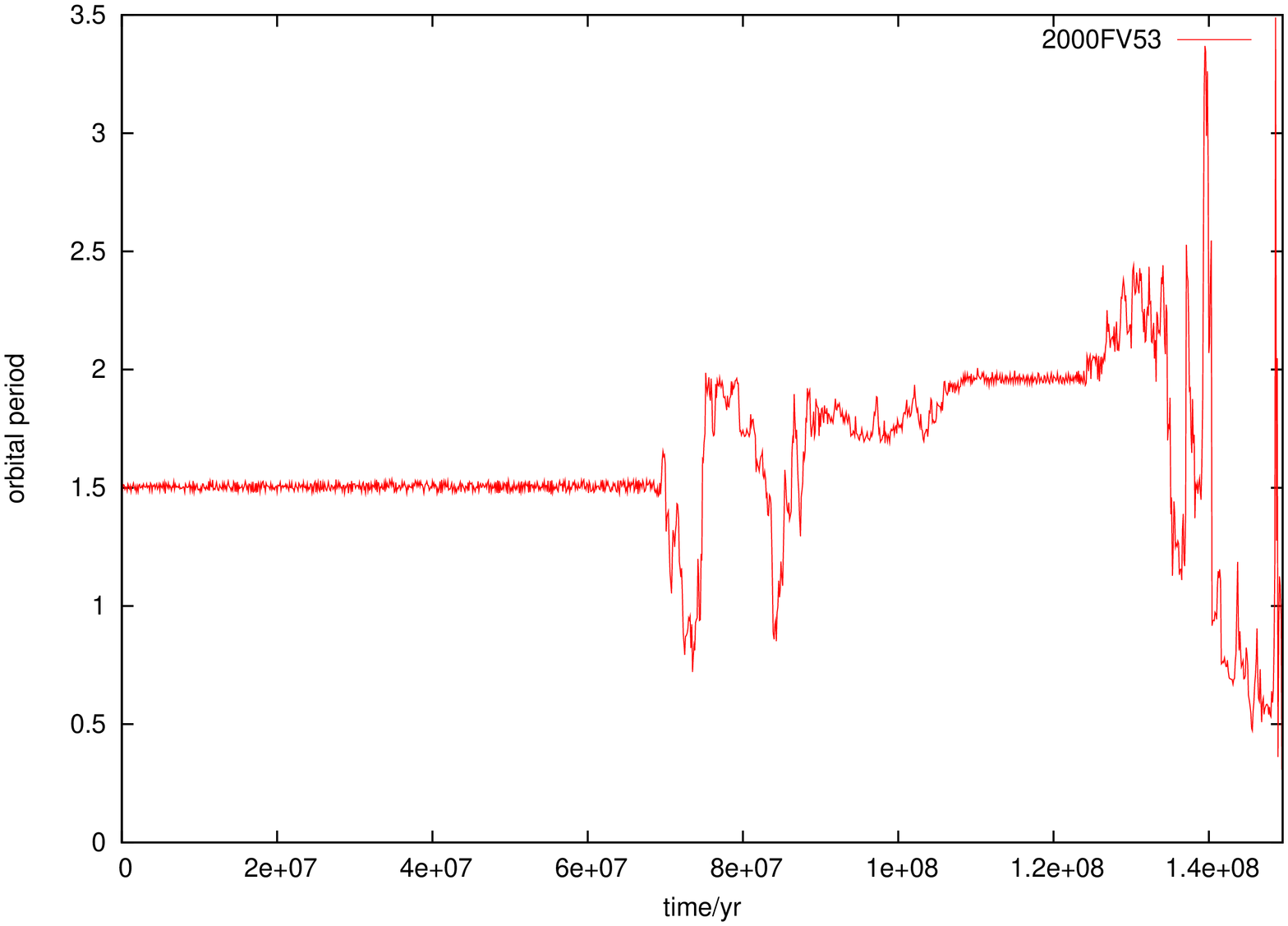}\\
\includegraphics[scale=0.35, angle=-90]{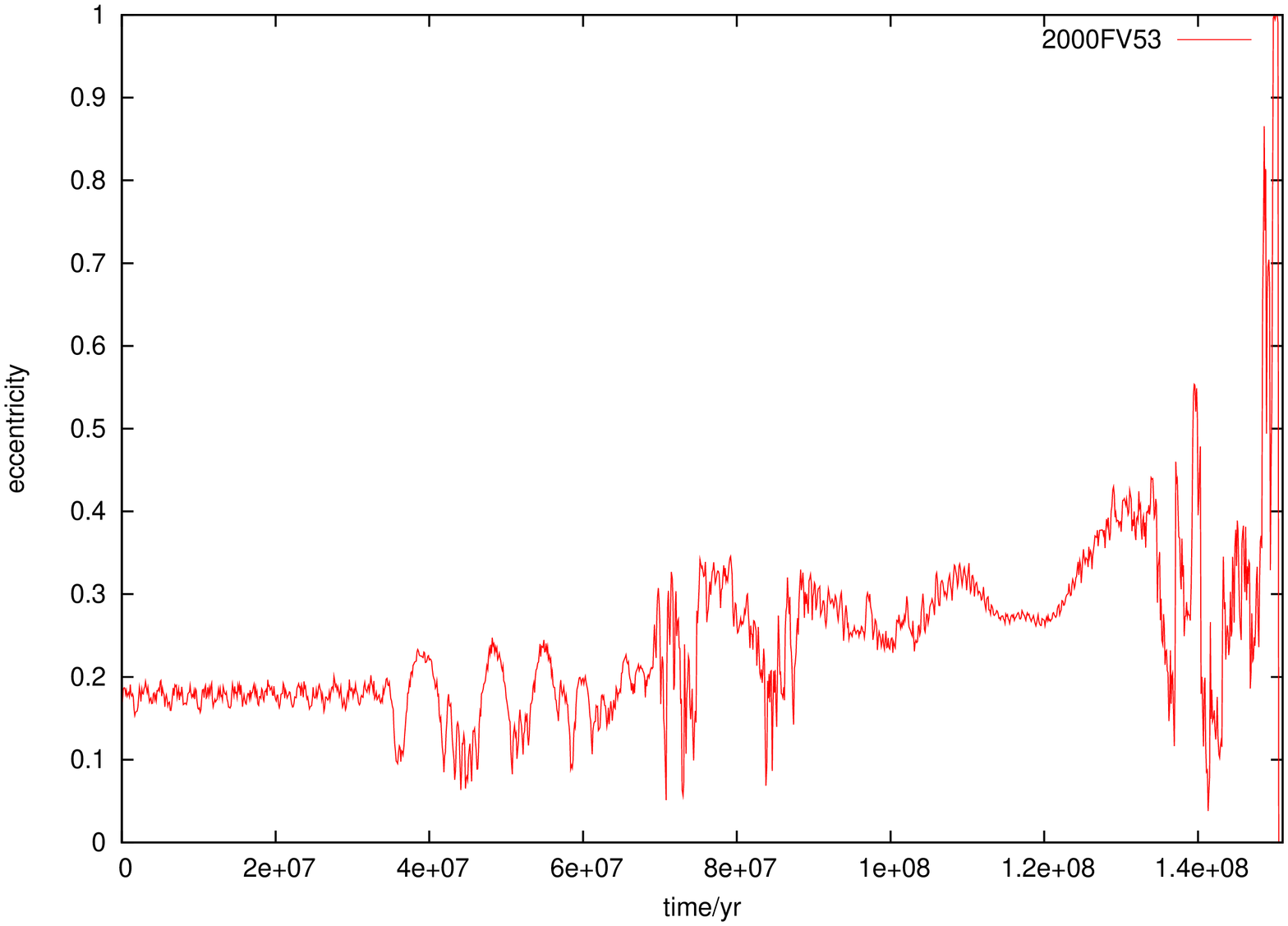}
\caption{Long-term evolution of the orbital period (over Neptune's orbital
period) and the eccentricity for Plutino 2000FV53 for 150~Myr.
Initially a Plutino, the orbit of this TNO is not stable and quits the 3:2
mean motion resonance just after about 69~Myr. Its eccentricity then increases
progressively and the TNO becomes a short-period comet that eventually
collides with a planet or the Sun. In this simulation it is ejected from the Solar
System after 150~Myr.} 
\label{2000FV53_orb_ecc}
\end{figure}

In Fig.\ref{2000FV53_orb_ecc} we plot the long-term evolution of the orbital
period and the eccentricity of TNO 2000FV53.
The initial 3:2 resonant configuration is abandoned just after 69~Myr, time after
which the eccentricity increases progressively. 
The TNO then undergoes close encounters with the planets, which will
increase the eccentricity even more, until it can reaches very high values.
The Plutino 2000FV53 then becomes a short-period comet and eventually collides
with a planet or the Sun. 
This mechanism has already been described as the responsible for the provision
of short period comets into the inner Solar System
\citep[e.g.][]{Morbidelli_1997}.
In our simulation the aforementioned TNO is ejected from the Solar System after
150~Myr. 

\subsection{Orbital overlap between Trojans and Plutinos}
\label{overlap}

\begin{figure*}[t!]
\begin{center}$
\begin{array}{cc cc}
\includegraphics[width=1.7in]{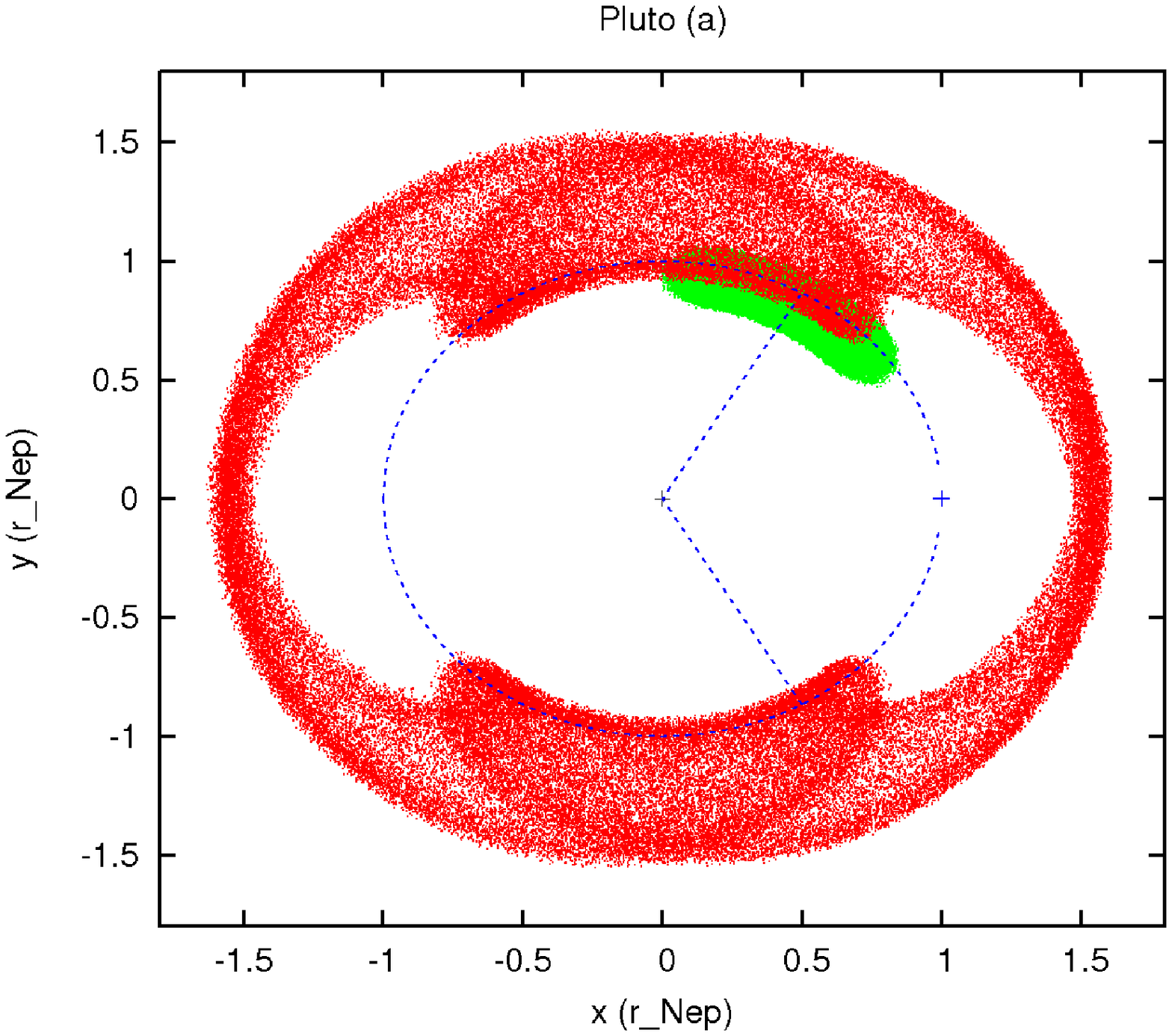} &
\includegraphics[width=1.7in]{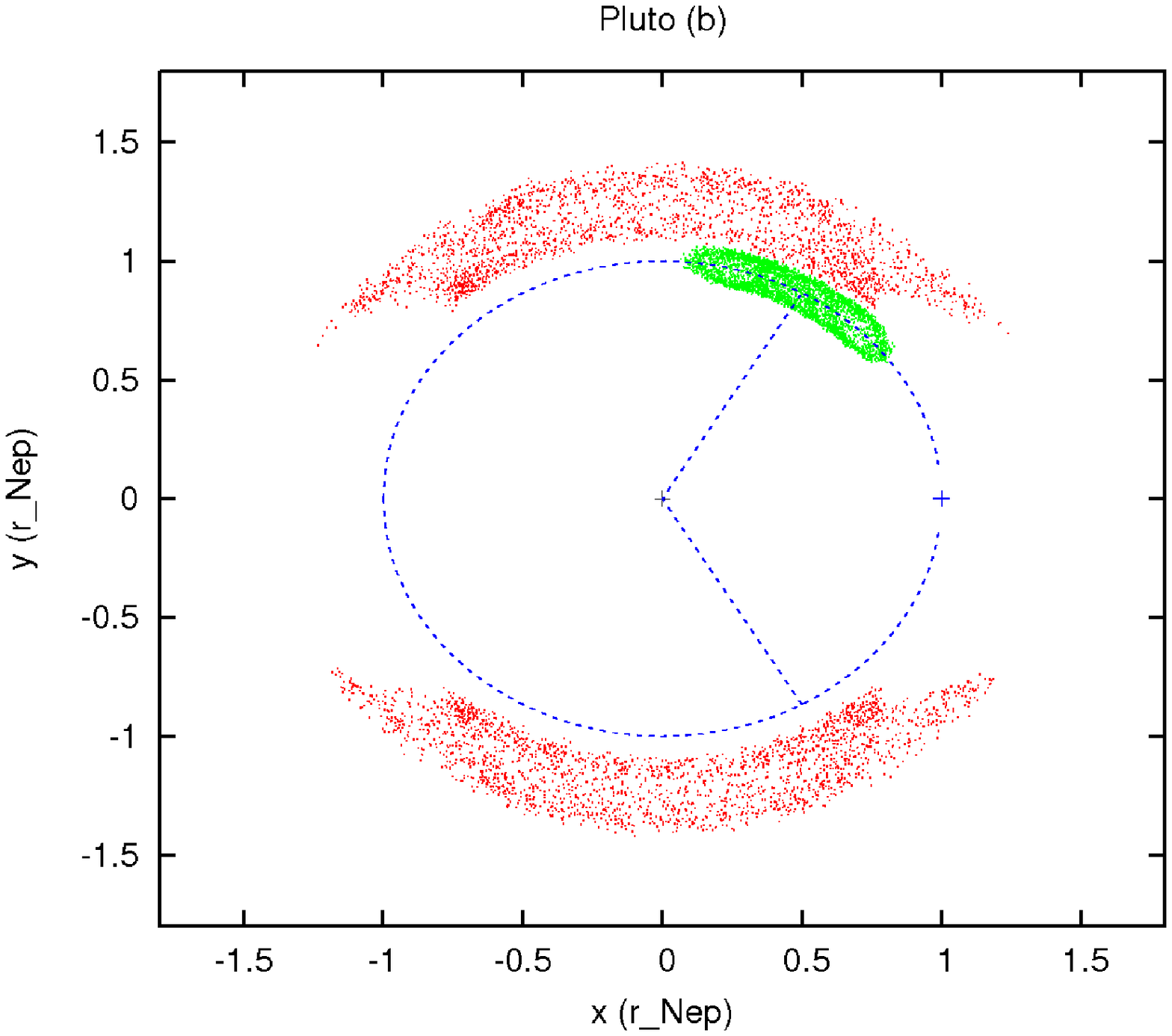} &
\includegraphics[width=1.7in]{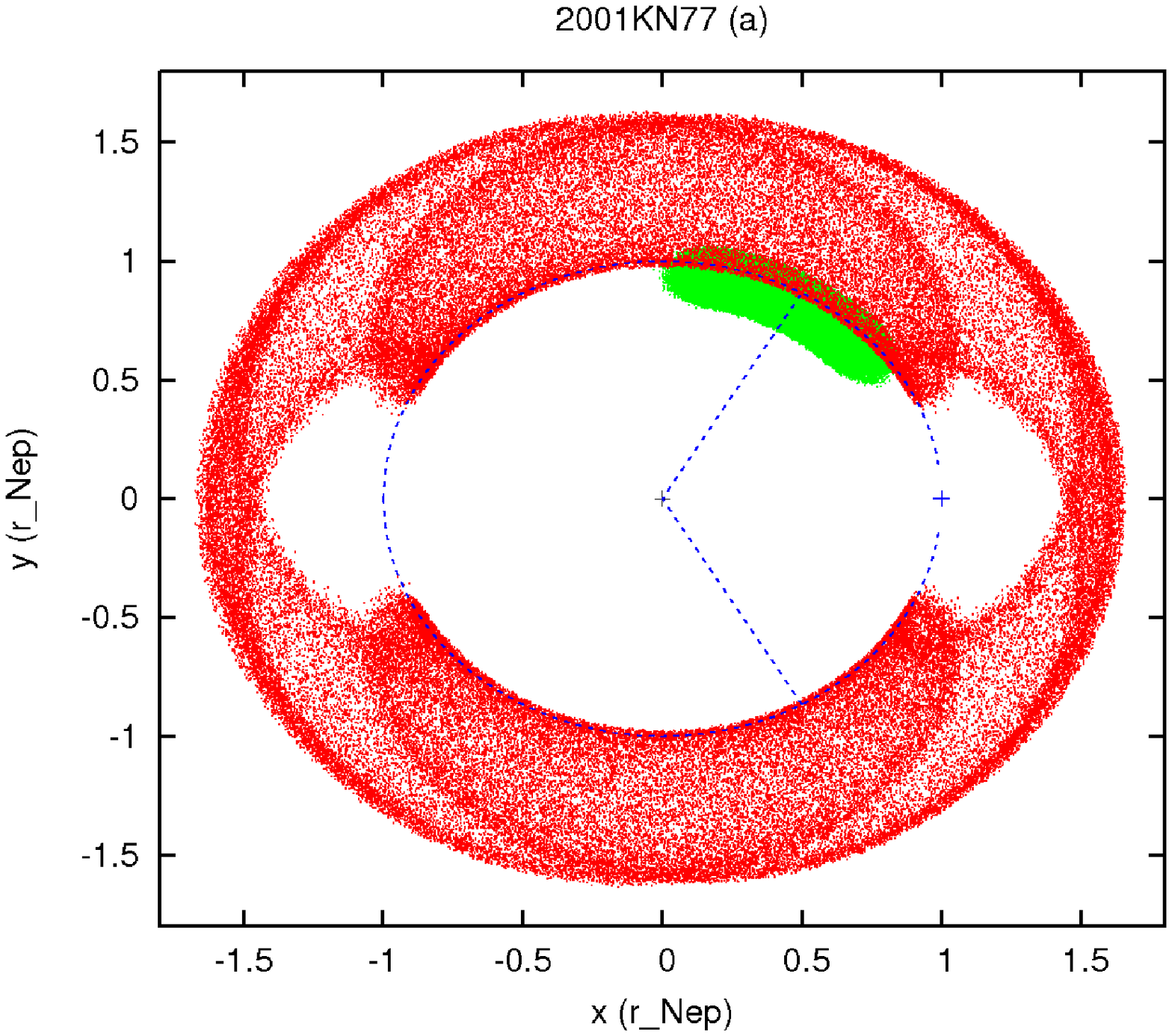} &
\includegraphics[width=1.7in]{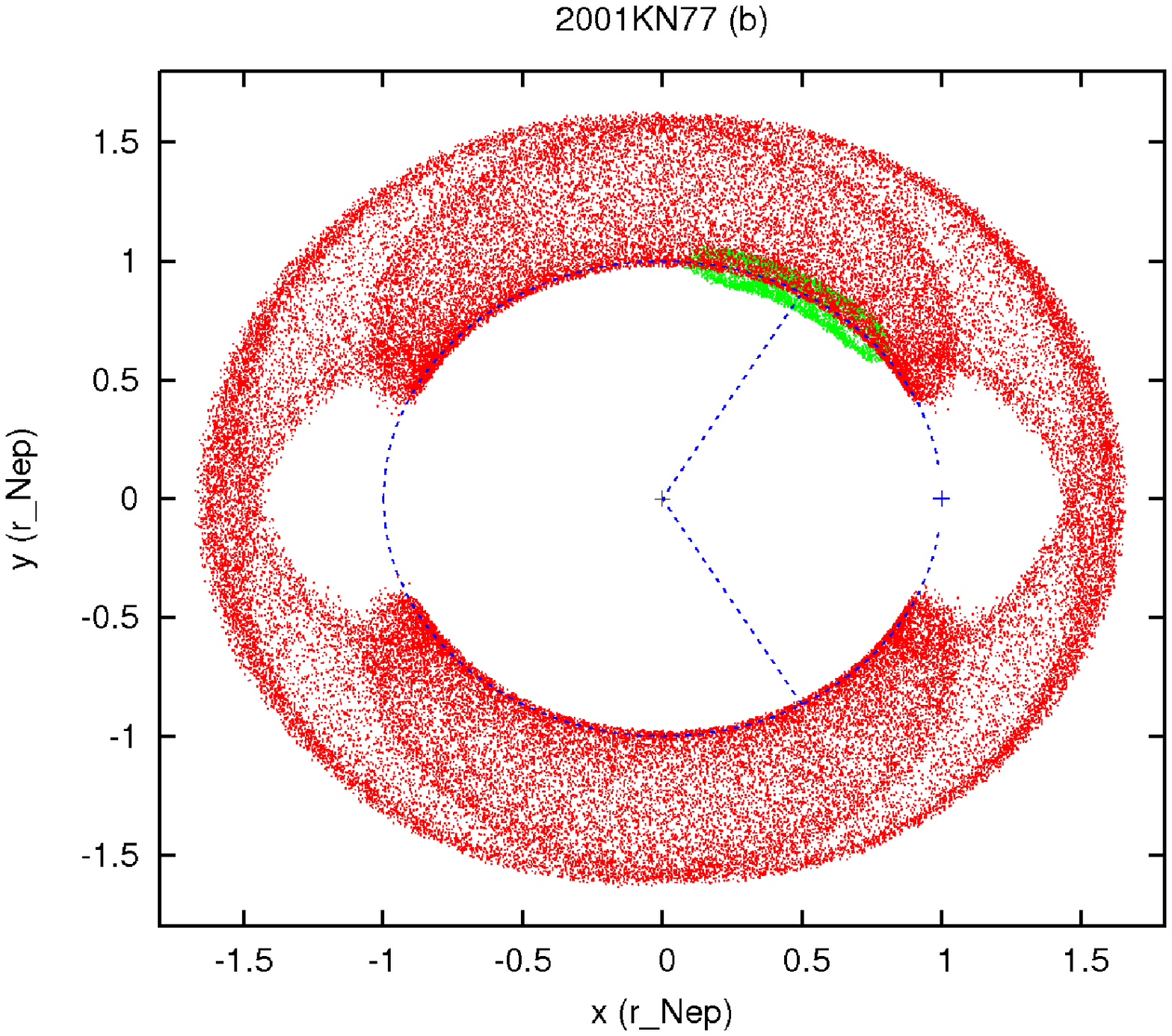} \\
\includegraphics[width=1.7in]{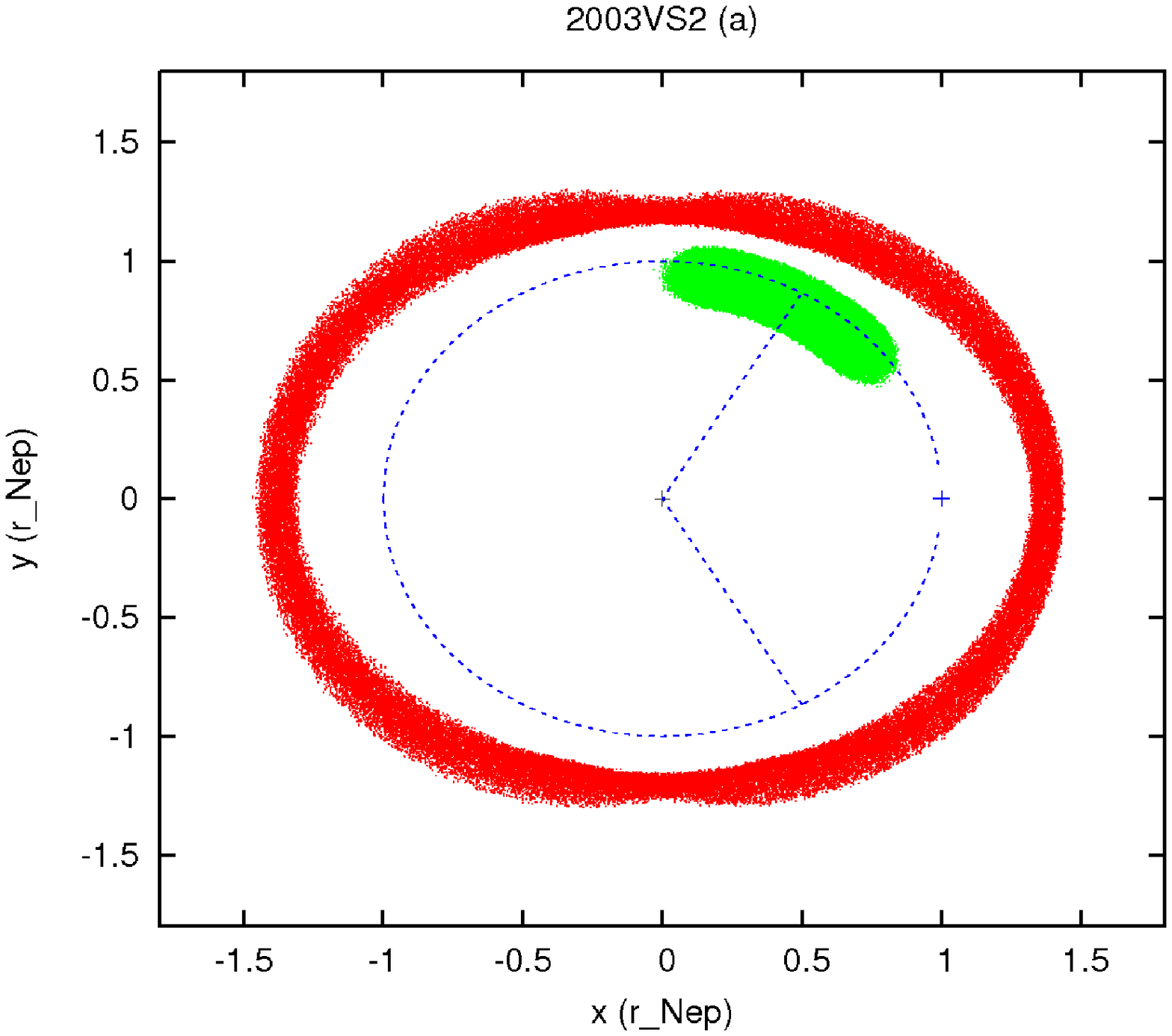} &
\includegraphics[width=1.7in]{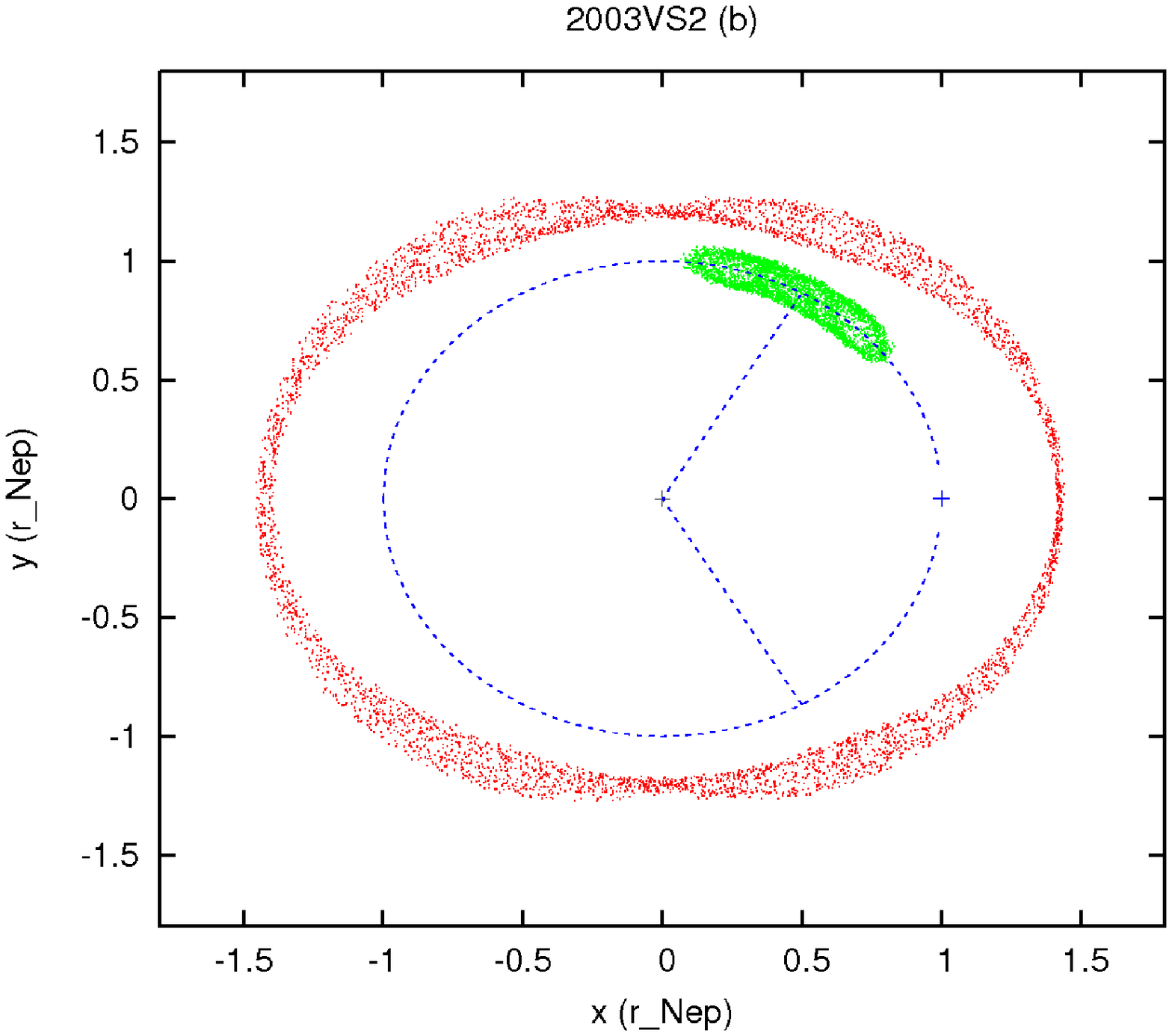} &
\includegraphics[width=1.7in]{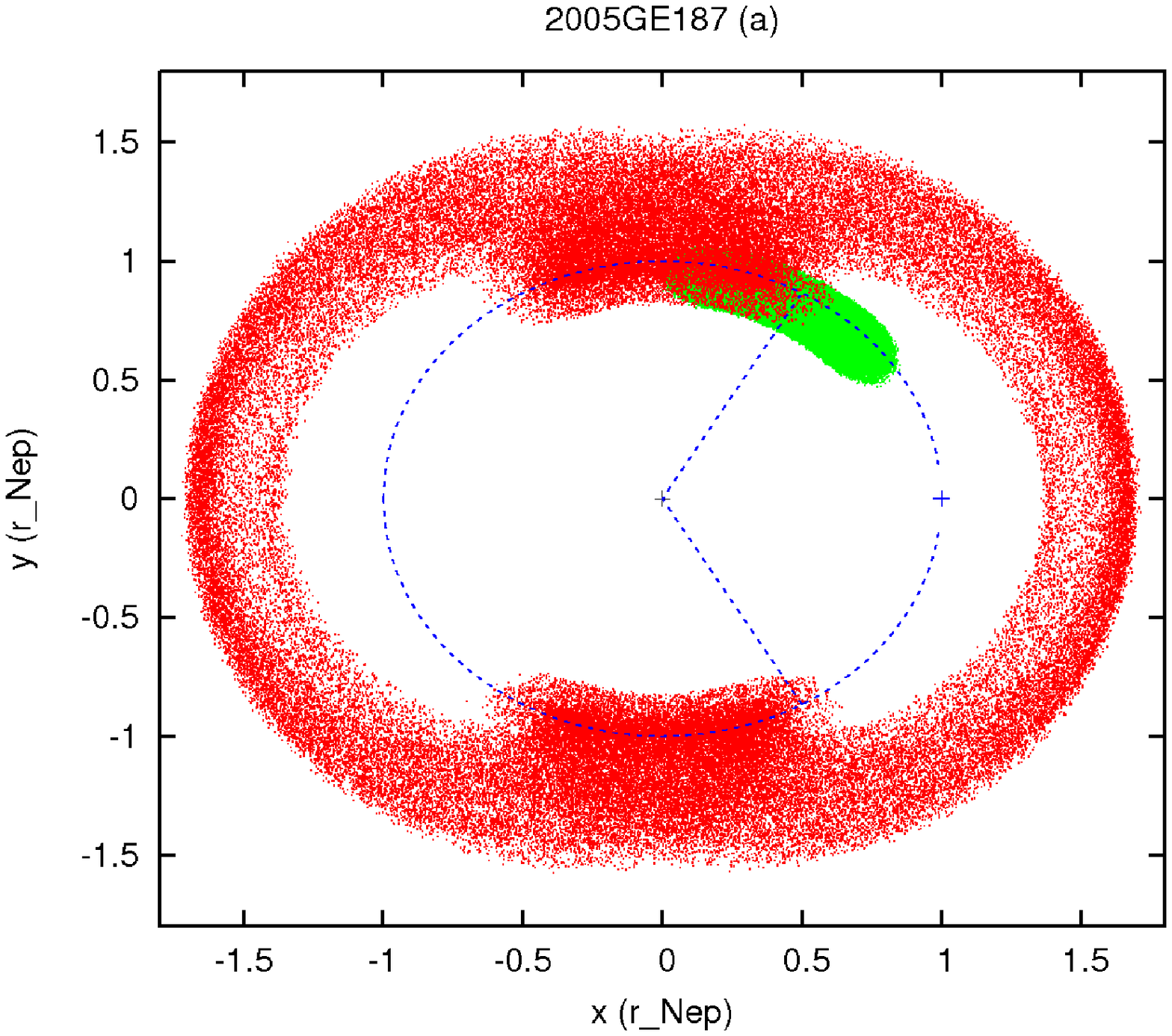} &
\includegraphics[width=1.7in]{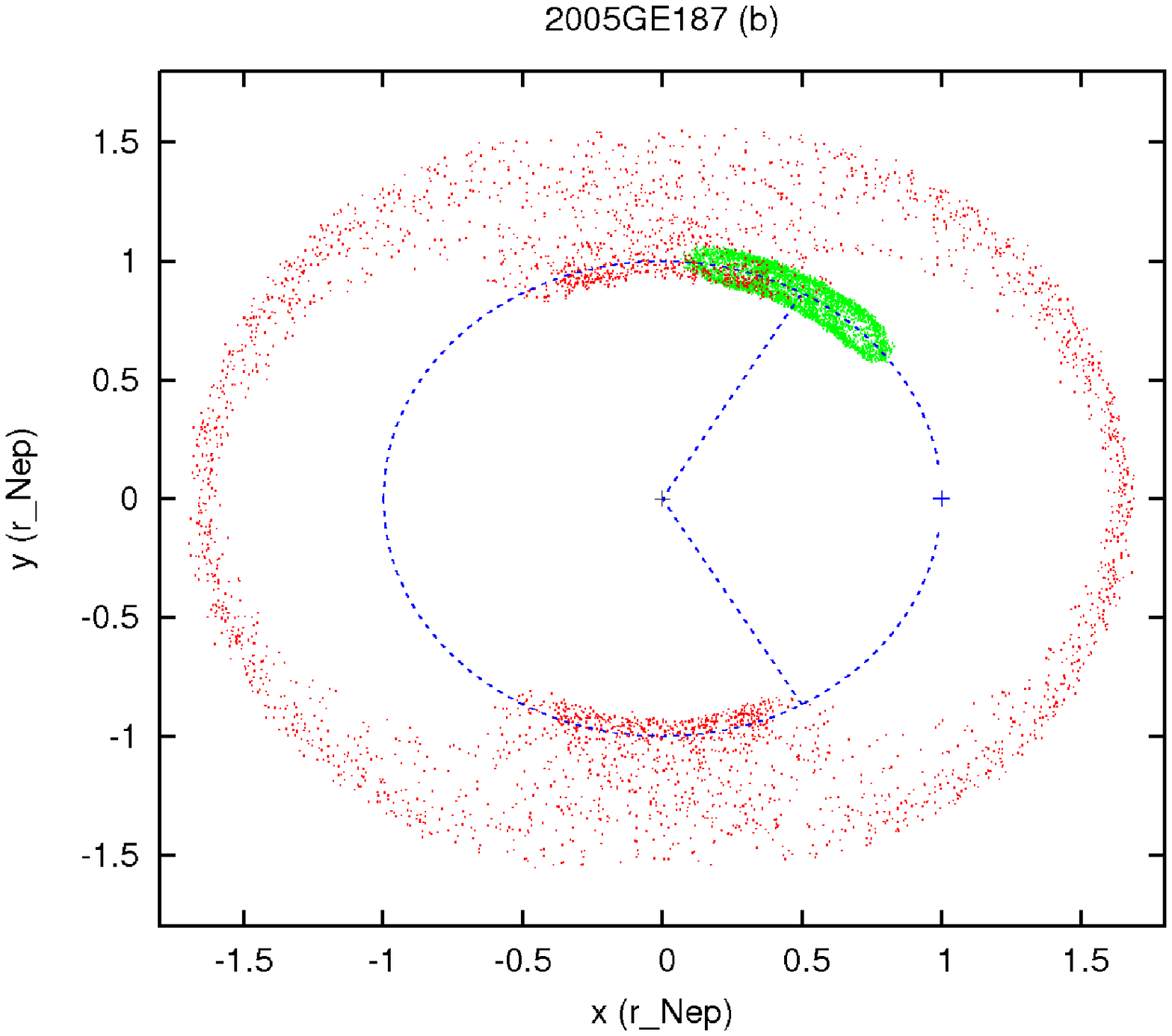} \\
\includegraphics[width=1.7in]{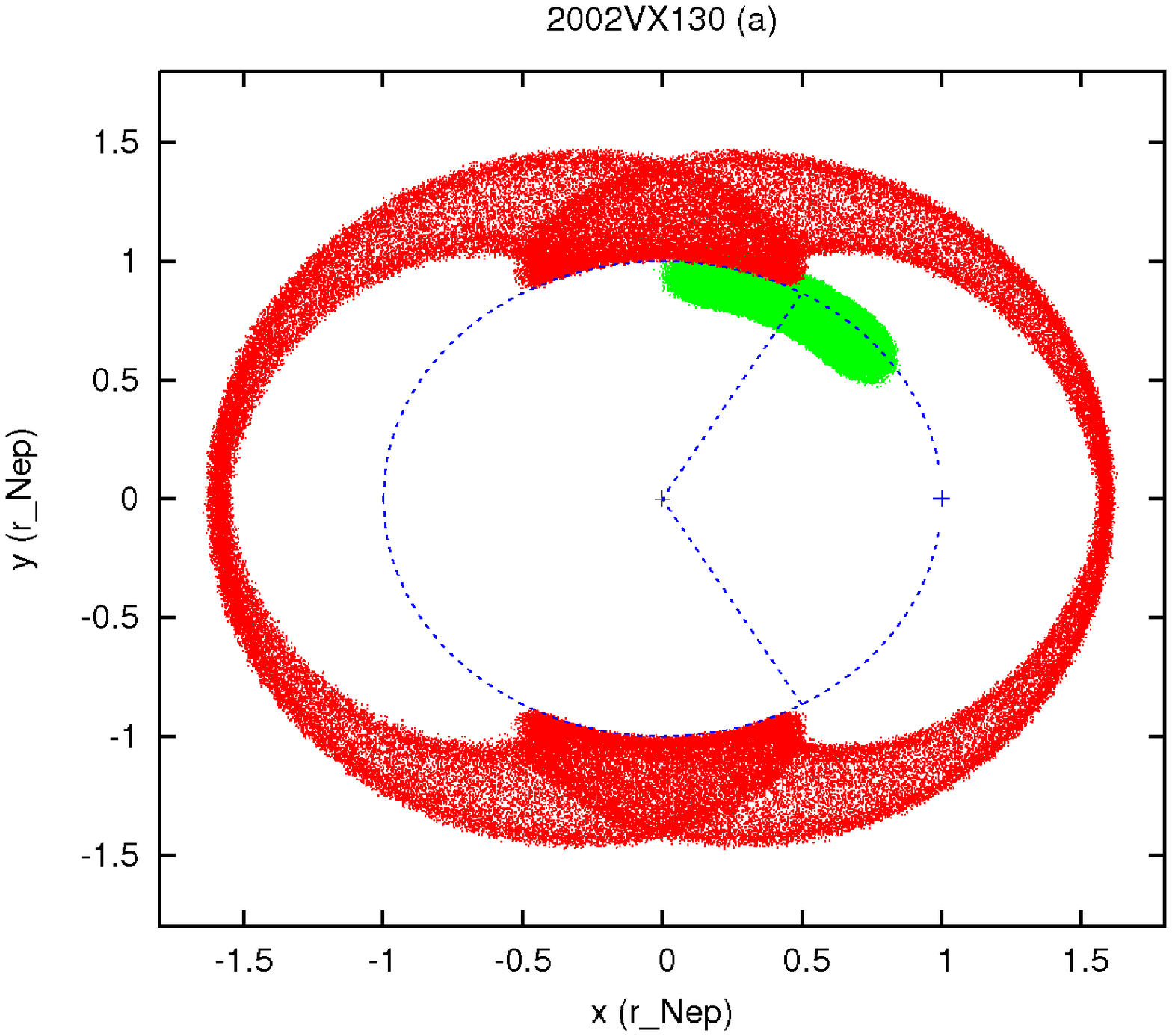} &
\includegraphics[width=1.7in]{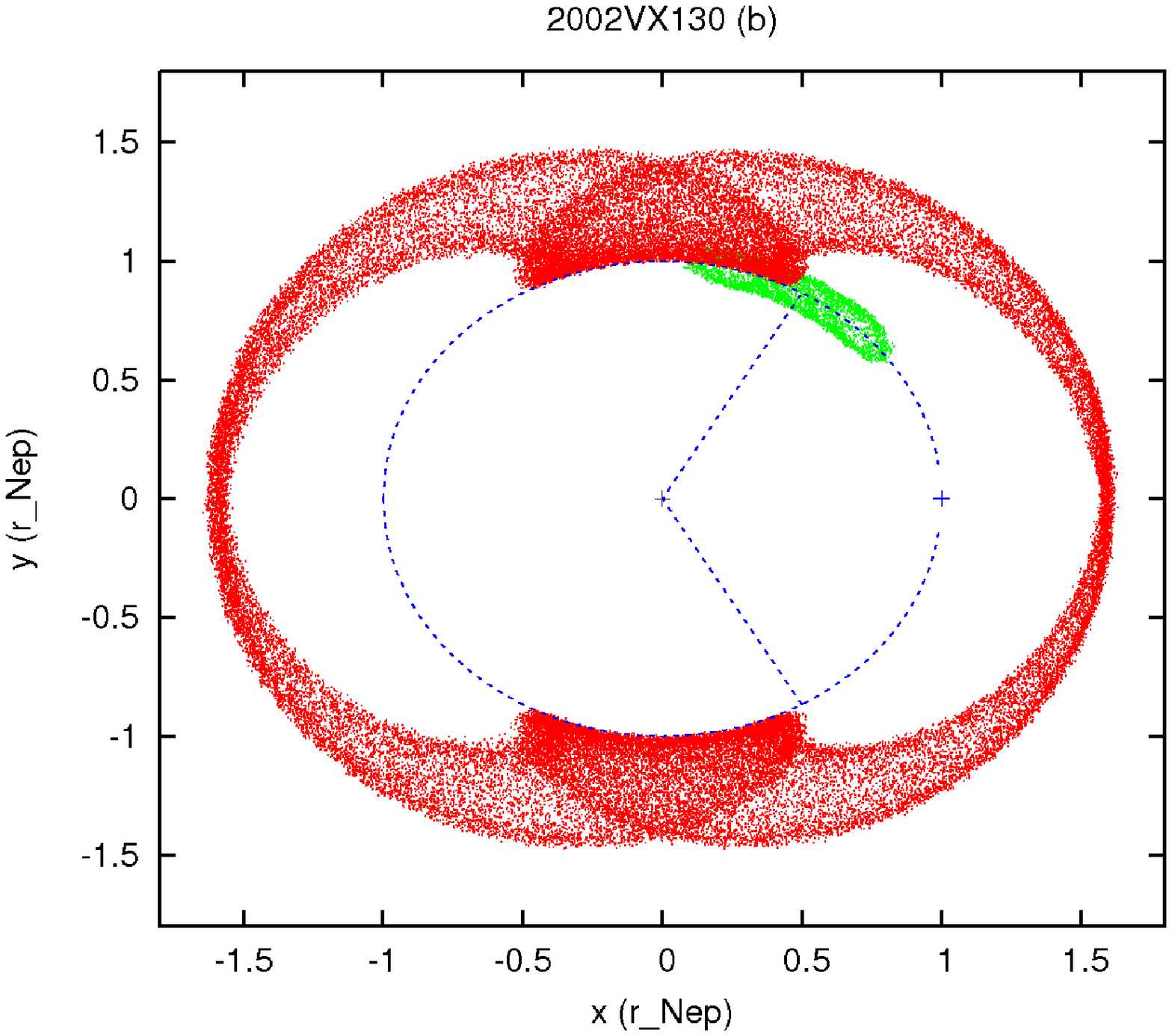} &
\includegraphics[width=1.7in]{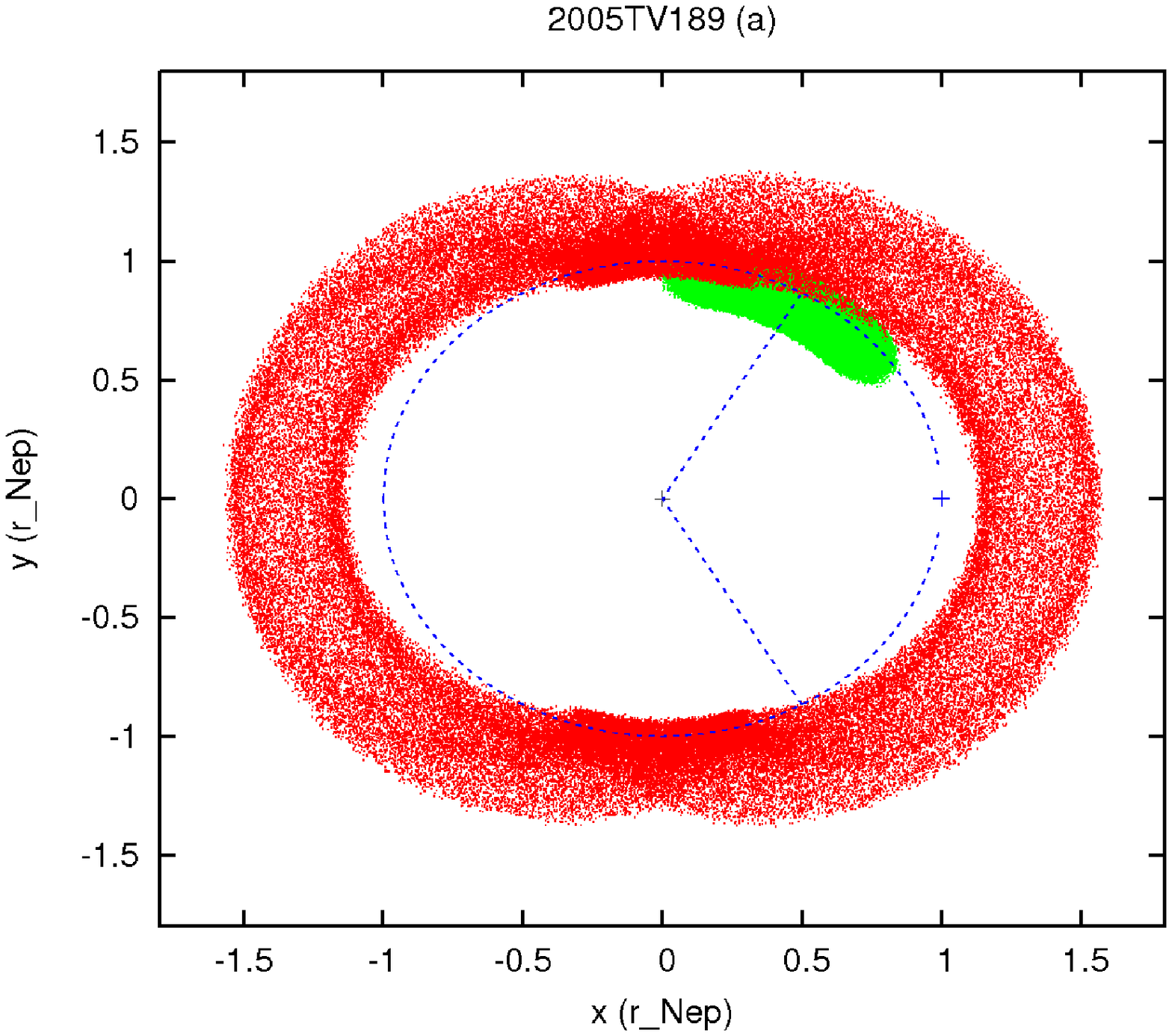} &
\includegraphics[width=1.7in]{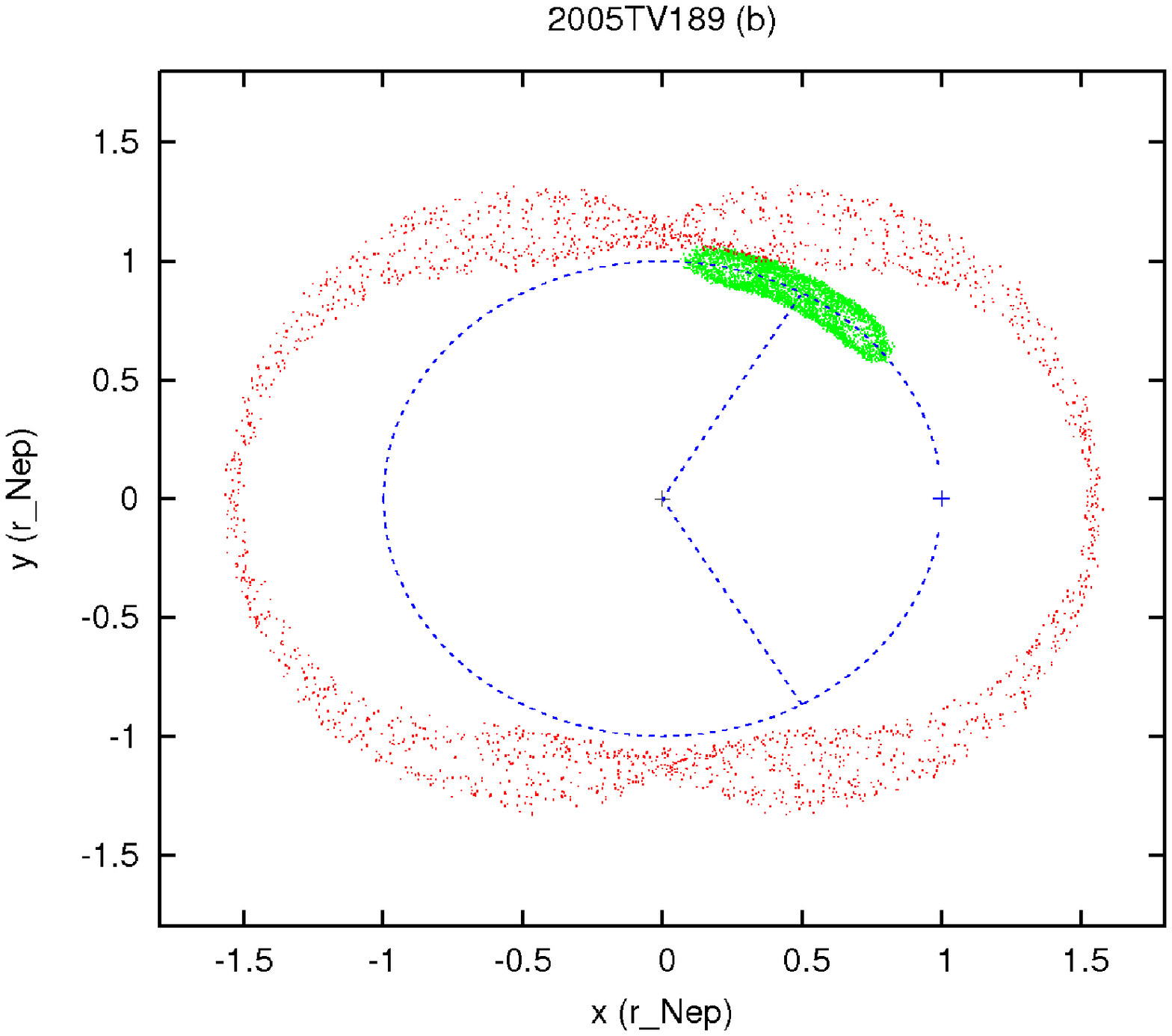}
\end{array}$
\end{center}
\caption{Orbital evolution of the Trojan 2007VL305 (in green) and Pluto plus most of the
Plutinos in Fig.\ref{plutinos} (in red), over 1~Gyr in the co-rotating frame of Neptune (in blue).
Left panel (a) shows the projection of the TNO position every 100~Kyr in the
orbital plane of Neptune, while right panels (b) shows the projection of the TNO
in the orbital plane of Neptune every 10~Kyr, only when the distance to this
plane is smaller then $10^{-3}$~AU. We observe that orbital overlap is
favored for Plutinos with high libration amplitudes, high eccentricity values
and low inclined orbits.} 
\label{superposition}
\end{figure*}

In Sect.\ref{Plutinos} we saw that because of libration, the Plutinos' orbits can
approach the Lagrangian point $L_4$.
In order to check the extent of the orbital overlap between Neptune Trojans and
Plutinos, in Fig.\ref{superposition} we plotted simultaneously their orbits in a
co-rotating frame with Neptune. 
We used the Trojan 2007VL305 for all representations, since it
has the most scattered orbit, maximizing the possibility of orbital merging with
a Plutino. 
For the Plutinos we used the same bodies as in
Sect.\ref{Plutinos}, which correspond to the extreme cases of
libration, eccentricity, and inclination.
The only exception is that Plutino 1996TP66 (with the smallest libration width) has
been replaced by Pluto, since a small libration amplitude does not allow the
Plutino to flyby the Lagrangian points.

Since the TNOs are not necessarily in
the same orbital plane as Neptune (especially for those having large inclination
values), two types of plots have been made: one where
we plotted the projection of the TNO position in the orbital plane of
Neptune (Fig.\ref{superposition}a), 
and another where we plotted the projection of the TNO in the
orbital plane of Neptune, only when the distance to this plane is smaller then
$10^{-3}$~AU (Fig.\ref{superposition}b), that is, less than 150\,000~km, half of the
Earth-Moon distance. 
Indeed, most of our Trojans have low inclinations and therefore lie close to
Neptune's orbital plane most of the time.
As a consequence, near this plane close encounters with Plutinos are
maximized.

The importance of plotting the two situations is clearly illustrated by
the behavior of Pluto (Fig.\ref{superposition}).
At first glance, looking only to the projection on Neptune's orbital plane, we
observe a large zone shared by the orbits of the two kinds of TNOs,
suggesting that close encounters may be a regular possibility.
However, when we restrain the plot only to Neptune's orbital plane, we observe
that Pluto is never on this plane when it approaches $L_4$, preventing any close
encounter with Trojans. 

From the analysis of Fig.\ref{superposition}, we can also conclude
that Plutinos with large libration amplitudes (e.g. 2001KN77) maximize the chances
of intercepting Trojans. 
This behavior was expected, as their orbits invade a large zone around the
Lagrangian point $L_4$.

On the other hand, Plutinos with low eccentricity values ($ e \sim 0.1 $; 
e.g. 2003VS2), always avoid Trojans independently of its
libration amplitude or orbital inclination, since the small eccentricity
prevents them from crossing Neptune's orbit. 
According to Fig.\ref{plutinopaths}, this result could also be expected, because for low
eccentricity there is no interception of the Plutino and Neptune's orbits.

Finally, when comparing the behavior of Plutinos in low-inclined orbits
(2001KN77 and 2002VX130) with high-inclined ones (Pluto, 2003VS2, 2005GE187 
and  2005TV189) we conclude that large orbital inclination
decreases the chances of close encounters because the Plutino is never close to
Neptune's orbital plane when it crosses the orbit of the planet.
The fact that Plutinos with low inclination share the Trojan space was
expectable, as both TNOs remain close to the same orbital plane.
However, Plutinos with high inclination could also approach the orbital plane of
Neptune at the moment they are close to the Lagrangian points, which is not
really observed.

From the above analysis (and also for the remaining 5 Trojans and 92 Plutinos),
we empirically conclude that orbital overlap between Trojans and 
Plutinos is favored for Plutinos with high libration amplitudes, high
eccentricity values and low inclined orbits.

\subsection{Collisions between Neptune Trojans and Plutinos}

In order to directly check if close encounters between the Neptune Trojans and
Plutinos can occur, and how often during 1~Gyr of numerical simulations, we
computed the distance between all bodies after each step-size.

For that purpose we arbitrarily selected two critical distances, one
$ d_1 < 2 \times 10^{-5}$~AU ($\sim$ 3\,000~km), for which we assume that the two
TNOs effectively collide, and a second $d_2 < 2 \times 10^{-3}$~AU ($\sim$
300\,000~km), for which the two bodies do not collide, but become closer
than the Earth-Moon distance.
Assuming a body with the mass density of water ice located at Neptune's 
distance to the Sun, $d_1$ and $d_2$ correspond approximately to the 
Hill radius of an object with diameter of 1.5~km and 150~km, respectivelly.
Alternatively, $d_2$ is equivalent to the Hill radius of a body with
0.01\% of Pluto's mass at the same distance from the Sun as Neptune.
TNOs with diameters of 150~km or larger when approaching at distances less than $d_2$
will be significantly perturbed by their mutual gravity and our model described in
Sect.\ref{dynamiceq} will no longer apply.
We assume that TNOs undergoing such close encounters may effectively
collide, or deviate considerably from their initial orbits and quit the resonant
configuration with Neptune. 

After 1~Gyr of simulations we did not observe any event for which the
minimal distance between two bodies, $d_{min} $, is lower than $d_1$, and
registered only 14 close encounters for which $ d_{min} < d_2 $.
The results are listed in Table\,\ref{tab:tabB}.
However, these results cannot be seen as definitive, but rather as minimal
estimations of close encounters.
Indeed, since our step-size is 0.1~yr, in a circular orbit a Trojan will travel
about 0.1~AU per step-size.
As a consequence, two TNOs may effectively collide between two step-sizes
and our program is unable to detect it.
The results listed in Table\,\ref{tab:tabB} must then be seen as indicative
of the possibility of collisions and not as conclusive.

\begin{table}[t!]
  \caption{Close encounters between TNOs during 1~Gyr ($ d_{min} < d_2 $). \label{tab:tabB}}  
  \begin{tabular}{c l l c r}
\hline
\hline 
\textbf{Type} & \textbf{Body 1} & \textbf{Body 2} & \textbf{time} & \textbf{$d_{min}$} \\ 
 &   &   & \textbf{(Gyr)} & \textbf{(km)} \\ \hline
T-P &2005TN53  &2002VU130 &0.244 &270\,226  \\
P-P &2003SO317 &1996SZ4   &0.278 &142\,977  \\
T-P &2005TO74  &1995HM5   &0.298 &227\,252  \\
P-P &1998HH151 &2001KX76  &0.320 &196\,351  \\
P-P &2001QH298 &1993SC    &0.406 &250\,561  \\
P-P &1998HH151 &2003AZ84  &0.455 &232\,759  \\
P-P &2001KB77  &2001QH298 &0.494 &281\,273  \\
P-P &2003SR317 &1994TB    &0.519 &271\,635  \\
P-P &2001KQ77  &2005TV189 &0.528 &193\,516  \\
P-P &1998WS31  &2004FW164 &0.542 &280\,527  \\
P-P &1998WV31  &1998HK151 &0.588 &236\,512  \\
T-T &2004UP10  &2005TO74  &0.849 &272\,851  \\
T-T &2004UP10  &2006RJ103 &0.864 &159\,886  \\
P-P &2004EH96  &1993SC    &0.986 &226\,779  \\
\hline
\end{tabular}
\end{table}

Assuming a constant speed for TNOs and a uniform distribution of their
relative minimal distances, 
we roughly estimate the real number of close encounters $ d_{min} < d_2 $ to be
50 times more frequent than those listed in Table\,\ref{tab:tabB} 
(0.1\,AU$/ d_2$ = 50).
Effective collisions ($ d_{min} < d_1 $) should also be more frequent in the
same proportion.
Thus, since $ d_1 = 10^{-2} d_2 $, the results showed in Table\,\ref{tab:tabB} for
$ d_{min} < d_2 $ can be seen as a rough indicator of the real number
of effective collisions occurring between Neptune Trojans and Plutinos.

Among the 14 ``collisions'' listed in Table\,\ref{tab:tabB}, two were between
Trojans, two between a Trojan and a Plutino, and the remaining ten between Plutinos.
It is not a surprise that Trojans or Plutinos also undergo close
encounters between each other,
because of the libration of their orbits (Fig.\,\ref{lib_plut}).
The two Plutinos that encounter Trojans are 1995HM5 ($e = 0.26$, $i =
4.8^\circ$, $\Delta \varphi = 70^\circ$) and 2002VU130 ($e = 0.21$, $i =
1.4^\circ$, $\Delta \varphi = 116^\circ$), both having small orbital
inclinations and large values for the eccentricity and libration amplitude.
This confirms our predictions from Sect.\ref{overlap}, that collisions between
Trojans and Plutinos are favored for Plutinos with high libration amplitudes,
high eccentricity values and low inclined orbits.
 
The fact that we count more close encounters between two Plutinos than between a
Trojan and a Plutino, cannot be seen as an indicator that
this last kind of encounter is less frequent.
Indeed, in our simulations the number of Trojans (6) is about 16 times smaller
than the number of Plutinos (98).
Therefore, there are roughly 16 times more chances of observing an
encounter between two Plutinos.
If our simulations had as much Trojans as it has Plutinos, we could then probably
expect to observe about 32 Trojan-Plutino encounters.
As a consequence, from the results listed in Table\,\ref{tab:tabB} we infer that
this kind of encounter is roughly 3 times more frequent than a Plutino-Plutino
one.
The same applies to the encounters between Trojans. In our simulation they are
also about 16 times less probable than a Trojan-Plutino encounter and we infer
that they should be about 50 times more frequent than encounters between
Plutinos.
Only by using a model with an identical initial number of Trojan and Plutinos,
would allow us to determine the exact proportions for each kind of encounter.

\section{Conclusions}
\label{conclusions}

In this work we aimed to verify if both Plutino and Neptune Trojan populations can be merged, and how frequently does that
occur, and examine if there could be a connection between such possibility and the observed properties 
of the two (sub-) populations of TNOs in question.
We analyzed the available colors and absolute magnitudes of Plutinos and Neptune Trojans. 
The nonexistence of significant albedo diversity among the large majority of these objects is assumed,
hence we interpreted absolute magnitudes as an equivalent to to object size. We find that:

\begin{description}
\item[(i)] there are no intrinsically bright (large) Plutinos at small inclinations;
\item[(ii)] there is an apparent excess of blue and intrinsically faint (small) Plutinos;
\item[(iii)] Neptune Trojans possess the same blue colors as Plutinos within the same (estimated) size
range do. 
\end{description}

From these results, we have hypothesized that there might have been some strong collisional interaction 
between Neptune Trojans and (1) the presently small Plutinos --- with the assumption that such collisions 
created only blue colored objects ---, or (2) the low inclined Plutinos --- with the assumption that such collisions 
created both blue and red objects. Paradoxically, the previous two opposite assumptions both seem to have 
observation support among other TNOs.

In order to differentiate between these two scenarios we performed a numerical simulation over 1~Gyr of the
future evolution of the outer Solar System composed by 5 planets, 6 Neptune Trojans, and 98 Plutinos.

By plotting simultaneously the orbits of the Neptune Trojans and Plutinos in a co-rotating frame with Neptune, it
becomes clear that there is a large overlap of the orbits of the two kinds of TNOs.
A more detailed analysis revealed, however, that close encounters with the Neptune Trojans
are favored for Plutinos with high libration amplitudes, high eccentricity values, and low inclined orbits. 
Tables \ref{tab:tab_3} and \ref{tab:tab_4} show the libration amplitudes and
periods computed for these objects.

After 1~Gyr of numerical simulations we registered 2 Trojan-Plutino, 2
Trojan-Trojan, and 10 Plutino-Plutino close approaches, i.e. less than $\sim$ 300\,000 km.
Since we have used the currently known objects, Trojans are much less numerous 
than Plutinos in our simulation. If the number of objects among each population is similar then we 
roughly infer that Trojan-Plutino and Trojan-Trojan ``collisions'' can be about three and
fifty times more frequent, respectively, than Plutino-Plutino collisions. 

The collision rates between Neptune Trojans and between Neptune Trojans and Plutinos might 
explain both why we observe a small number of Neptune Trojans and why there is an absence of large
Neptune Trojans: a strong collisional evolution possibly played an important role by shattering
and/or depleting Neptune Trojans. 

Our results also show that Plutinos in
low-inclined orbits have more chances of colliding
with Neptune Trojans. This result gives no support for a (mutual) collisional
origin of the equal-sized and equal-colored Neptune Trojans and small Plutinos,
as the latter are equally spread in inclination (see Sect. \ref{obs}, scenario~\#1). On the other hand, it
gives plausibility to the origin of the concentration of small Plutinos with 
$i<8-13^\circ$ as being a consequence of
some collisional interaction with Neptune Trojans(see Sect. \ref{obs}, scenario~\#2). 

During our numerical simulations we also observed that the orbit of Trojan 2001QR322 became unstable as
well as the orbits of 10 more Plutinos. The stability of Neptune Trojans appears to be enhanced for high inclination
values \citep{2007MNRAS.382.1324D}, and also for small libration amplitudes.
On the other hand, Plutinos seem to be essentially pushed out of their resonance by
Pluto, in conformity with the results of \citet{1999AJ....118.1873Y} and \citet{Nesvorny_etal_2000}.

This work's results were derived for objects considered as test particles (except for Pluto) 
and the number of ``collisions'' were extrapolated from the amount of close encounters after each
integration step-size. Accurate estimations for the amount of collisions between Neptune Trojans 
and Plutinos can only be made with the inclusion of mutual gravitational interactions between all objects,  
and using the real spatial and size distributions of these objects. 

Nonetheless, our sketchy analysis indicates that under certain assumptions, 
which have parallel in what has been inferred for other TNOs, shattering collisions involving 
Neptune Trojans might have played a crucial role on the creation of the size-inclination asymmetries 
observed among Plutinos. We recall that we have disregarded the possibility that these asymmetries 
might have been created by collisional interaction with some other sub-population of TNOs than the 
Neptune Trojans in view of the results obtained by \cite{2003Icar..162...27T} and \cite{2003EM&P...92..233T}. 
If those works come to be revealed as inadequate approximations of the relative collision rates suffered by 
TNOs our inference on a possible cause for the size-inclination distribution of Plutinos looses its ground.  
Further, if Neptune Trojans and low inclined Plutinos do not possess identical spin rate {\it versus} size distributions, 
which should be distinct from the higher inclined Plutinos, then our suggestions cannot hold either \citep[e.g.][]{1981A&A...104..159F}.  
More detailed studies on the interaction between Neptune Trojans and Plutinos should be attempted.

\section*{Acknowledgments}

The authors thank the referee D. Nesvorn\'y for his comments that helped to improve this document and to 
M. H. M. Morais for discussions.
This work was supported by the Funda\c{c}\~{a}o para a Ci\^{e}ncia e a
Tecnologia (Portugal).

\appendix

\section{Additional Tables}


\longtab{1}{
\begin{longtable}{|c|c|c|c|c|}
\caption{\small Data relative to Trojans and Plutinos.} \label{tab:tab_1}\\
\hline

\textbf{Name} & \textbf{Sample} & \textbf{Class$^a$} & \textbf{B-R$^b$} & \textbf{$H_R$$^c$} \\
\hline
\endfirsthead

\caption{continued.}\\ 
\hline
\textbf{Name} & \textbf{Sample} & \textbf{Class$^a$} & \textbf{B-R$^b$} & \textbf{$H_R$$^c$} \\
\hline
\endhead

\multicolumn{5}{|r|}{\small\sl continued on next page} \\ \hline
\endfoot

\hline
\endlastfoot

\hline
2001QR322 &ST06 &1:1 &1.26 &7.67  \\ \hline
2004UP10  &ST06 &1:1 &1.16 &8.50  \\ \hline
2005TN53  &ST06 &1:1 &1.29 &8.89  \\ \hline
2005TO74  &ST06 &1:1 &1.34 &8.29  \\ \hline
Pluto     &JL01   &3:2 &1.34 &-1.37  \\ \hline
1993RO    &TR00   &3:2 &1.36 &8.41   \\ \hline
1993SB    &TR00   &3:2 &1.29 &7.68   \\ \hline
1993SC    &TR98/RT99   &3:2 &1.97 &6.53   \\ \hline
1994JR1   &M2S99   &3:2 &1.61 &7.06   \\ \hline
1994TB    &TR98/RT99   &3:2 &1.78 &7.43   \\ \hline
1995HM5   &TR98/RT99   &3:2 &1.01 &7.88   \\ \hline
1995QY9   &M2S99   &3:2 &1.21 &7.02   \\ \hline
1995QZ9   &TR00   &3:2 &1.40 &8.06   \\ \hline
1996RR20  &TR00   &3:2 &1.87 &6.49   \\ \hline
1996SZ4   &TR00   &3:2 &1.35 &7.92   \\ \hline
1996TP66  &TR98/RT99   &3:2 &1.85 &6.71   \\ \hline
1996TQ66  &TR98/RT99   &3:2 &1.86 &6.99   \\ \hline
1997QJ4   &LP02  &3:2 &1.10 &7.84   \\ \hline
1998HK151 &M2S02   &3:2 &1.24 &6.78   \\ \hline
1998UR43  &MBOSS    &3:2 &1.35 &8.09   \\ \hline
1998US43  &LP04  &3:2 &1.19 &7.75   \\ \hline
1998VG44  &TRC07   &3:2 &1.52 &6.10   \\ \hline
1998WS31  &LP04  &3:2 &1.31 &7.77   \\ \hline
1998WU31  &LP04  &3:2 &1.23 &7.99   \\ \hline
1998WV31  &LP04  &3:2 &1.34 &7.53   \\ \hline
1998WW24  &LP04  &3:2 &1.35 &7.84   \\ \hline
1998WZ31  &LP04  &3:2 &1.26 &7.93   \\ \hline
1999TC36  &TRC03   &3:2 &1.74 &4.64   \\ \hline
1999TR11  &TR00   &3:2 &1.77 &7.88   \\ \hline
2000EB173 &TR03   &3:2 &1.60 &4.43   \\ \hline
2000GN171 &TRC07   &3:2 &1.57 &5.62   \\ \hline
2001KB77  &TRC07   &3:2 &1.39 &7.18   \\ \hline
2001KD77  &LP04  &3:2 &1.75 &5.74   \\ \hline
2001KX76  &M2S02   &3:2 &1.64 &3.25   \\ \hline
2001KY76  &M2S05   &3:2 &1.85 &6.68   \\ \hline
2001QF298 &TRC07   &3:2 &1.14 &4.91   \\ \hline
2002GF32  &M2S05   &3:2 &1.76 &5.95   \\ \hline
2002GV32  &M2S05   &3:2 &1.96 &6.75   \\ \hline
2002VE95  &TRC07   &3:2 &1.79 &5.06   \\ \hline
2002VR128 &TRC07   &3:2 &1.54 &4.83   \\ \hline
2002XV93  &TRC07   &3:2 &1.09 &4.36   \\ \hline
2003AZ84  &TRC07   &3:2 &1.06 &3.46   \\ \hline
2003VS2   &TRC07   &3:2 &1.52 &4.14   \\ \hline
2004DW    &TRC07   &3:2 &1.05 &1.92   \\ \hline
2004EW95  &TRC07   &3:2 &1.08 &6.08   \\ \hline
\end{longtable}
$^a$ {Resonance with Neptune.}
$^b$ {Color index.}
$^c$ {R-filter Absolute Magnitude.}
Samples -- [TR98/RT99]: {\cite{1998Natur.392...49T}, \cite{1999Natur.398..129R}};
[TR00]: {\cite{2000Natur.407..979T}};
[TR03]: {\cite{2003Icar..161..181T}};
[TRC03]: {\cite{2003ApJ...599L..49T}};
[TRC07]: {Database: http://www.physics.nau.edu/~tegler/research/survey.htm};
[LP02]: {\cite{2002A&A...395..297B}}; 
[LP04]: {\cite{2004Icar..170..153P}};
[MBOSS]: {Database: \cite{2002A&A...389..641H}, \cite{2001A&A...380..347D}, \cite{2001Icar..152..246G}};
[M2S99]: {\cite{1999Icar..142..476B}};
[M2S02]: {\cite{2002AJ....124.2279D}};
[M2S05]: {\cite{2005Icar..174...90D}};
[JL01]: {\cite{2001AJ....122.2099J}};
[ST06]: {\cite{2006Sci...313..511S}}.
}


\begin{table*}
\centering
  \caption{\small{Orbital data for the Giant Planets and Pluto at JD\,2454200.50 (http://ssd.jpl.nasa.gov/horizons.cgi).}} \label{tab:tab_2}
  \begin{tabular}{|c|c|c|c|c|c|c|c|}
\multicolumn{8}{c}%
{{\bfseries}}\\
\hline \multicolumn{1}{|c|}{\textbf{Name}} & \multicolumn{1}{c|}{\textbf{$a$ (AU)}} & \multicolumn{1}{c|}{\textbf{$e$}} & \multicolumn{1}{c|}{\textbf{$i$ (deg)}} & \multicolumn{1}{c|}{\textbf{$M$ (deg)}} & \textbf{$\omega$ (deg)} & \textbf{$\Omega$ (deg)} & \textbf{$m$ (M$_\odot$)} \\
\hline
Jupiter &5.20219308  &0.04891224 &1.30376425 &240.35086842 &274.15634048 &100.50994468 &9.5479194$\times 10^{-4}$ \\ \hline
Saturn  &9.54531447  &0.05409072 &2.48750693 &45.76754755  &339.60245769 &113.63306105 &2.8586434$\times 10^{-4}$ \\ \hline
Uranus  &19.19247127 &0.04723911 &0.77193683 &171.41809349 &98.79773610  &73.98592654  &4.3558485$\times 10^{-5}$ \\ \hline
Neptune &30.13430686 &0.00734566 &1.77045595 &293.26102612 &255.50375800 &131.78208581 &5.1681860$\times 10^{-5}$ \\ \hline
Pluto   &39.80661969 &0.25440229 &17.121129  &24.680638    &114.393972   &110.324800   &6.5607561$\times 10^{-9}$ \\ \hline
\end{tabular}
\end{table*}


\begin{table*}
\centering
  \caption{\small{Orbital data for the Neptune Trojans at JD\,2454200.50 (ftp://ftp.lowell.edu/pub/elgb/astorb.html).}} \label{tab:tab_3}
  \begin{tabular}{|c|c|c|c|c|c|c|c|c|c|c|}
\multicolumn{8}{c}%
{{\bfseries}}\\
\hline 
\textbf{$\#$} & \textbf{Name} & 
\textbf{$a$ (AU)}  & \textbf{$e$} & \textbf{$i$ (deg)} &
\textbf{$M$ (deg)} & \textbf{$\omega$ (deg)}  & \textbf{$\Omega$ (deg)} & 
\textbf{$P_{lib}$ (Kyr)} & \textbf{$\varphi_0$ (deg)} & \textbf{$\Delta\varphi$ (deg)}\\  
\hline
1 & 2001QR322  &30.190 &0.029 &1.3  &60.2  &154.8 &151.7 &9.23 &68.10 &25.90 \\ \hline
2 & 2004UP10   &30.099 &0.025 &1.4  &334.1 &2.2   &34.8  &8.86 &61.44 &10.5  \\ \hline 
3 & 2005TN53   &30.070 &0.062 &25.0 &280.3 &88.6  &9.3   &9.42 &58.95 &6.61  \\ \hline
4 & 2005TO74   &30.078 &0.051 &5.3  &260.1 &306.9 &169.4 &8.80 &60.91 &6.88  \\ \hline
5 & 2006RJ103  &29.973 &0.028 &8.2  &226.6 &35.4  &120.8 &8.87 &60.45 &6.13  \\ \hline
6 & 2007VL305  &29.956 &0.061 &28.1 &348.5 &216.1 &188.6 &9.57 &61.08 &14.26 \\ \hline
\end{tabular}
\end{table*}


\longtab{4}{
\begin{longtable}{|c|c|c|c|c|c|c|c|c|c|c|}

\caption{\small{Orbital data for the Plutinos at JD\,2454200.50 (ftp://ftp.lowell.edu/pub/elgb/astorb.html).}} \label{tab:tab_4}\\

\multicolumn{11}{c}%
{{\bfseries}}\\
\hline 
\textbf{$\#$} & \textbf{Name} & 
\textbf{$a$ (AU)}  & \textbf{$e$} & \textbf{$i$ (deg)} &
\textbf{$M$ (deg)} & \textbf{$\omega$ (deg)}  & \textbf{$\Omega$ (deg)} & 
\textbf{$P_{lib}$ (Kyr)} & \textbf{$\varphi_0$ (deg)} & \textbf{$\Delta\varphi$ (deg)}\\  
\hline
\endfirsthead

\multicolumn{11}{c}%
{{\tablename\ \thetable{} -- Continuation from previous page}} \\
\hline 
\textbf{$\#$} & \textbf{Name} & 
\textbf{$a$ (AU)}  & \textbf{$e$} & \textbf{$i$ (deg)} &
\textbf{$M$ (deg)} & \textbf{$\omega$ (deg)}  & \textbf{$\Omega$ (deg)} & 
\textbf{$P_{lib}$ (Kyr)} & \textbf{$\varphi_0$ (deg)} & \textbf{$\Delta\varphi$ (deg)}\\  
\hline
\endhead

\multicolumn{11}{|r|}{{Continues in next page}} \\ \hline
\endfoot

\hline
\endlastfoot

\hline
 1  &1993RO     &39.118  &0.196  &3.717  &14.487    &187.832   &170.337 &16.63 &178.23 &113.94\\ \hline
 2  &1993SB     &39.171  &0.317  &1.939  &336.923   &79.282    &354.837 &20.45 &179.78 &53.19\\ \hline
 3  &1993SC     &39.438  &0.186  &5.161  &53.290    &316.131   &354.662 &20.15 &178.98 &72.12\\ \hline
 4  &1994JR1    &39.631  &0.123  &3.803  &15.562    &102.750   &144.734 &19.72 &-177.19 &86.60\\ \hline
 5  &1994TB     &39.329  &0.314  &12.136 &342.780   &99.006    &317.365 &21.15 &178.82 &46.23\\ \hline
 6  &1995HM5    &39.842  &0.258  &4.809  &340.199   &59.756    &186.637 &19.91 &-178.97 &69.58\\ \hline
 7  &1995QY9    &39.586  &0.262  &4.837  &1.472     &24.792    &342.061 &15.39 &178.82 &120.22\\ \hline
 8  &1995QZ9    &39.329  &0.145  &19.580 &47.100    &141.846   &188.035 &21.90 &178.76 &16.37\\ \hline
 9  &1996RR20   &39.522  &0.177  &5.311  &128.591   &48.888    &163.546 &20.33 &182.44 &69.37\\ \hline
10  &1996SZ4    &39.422  &0.255  &4.743  &354.409   &30.010    &15.977  &18.71 &179.04 &90.02\\ \hline
11  &1996TP66   &39.209  &0.328  &5.693  &10.283    &75.084    &316.736 &21.51 &180.08 &7.15\\ \hline
12  &1996TQ66   &39.263  &0.119  &14.680 &12.619    &18.946    &10.769  &23.46 &172.12 &10.16\\ \hline
13  &1997QJ4    &39.251  &0.224  &16.575 &324.580   &82.174    &346.843 &20.50 &179.81 &72.94\\ \hline
14  &1998HH151  &39.640  &0.194  &8.774  &349.887   &33.586    &194.779 &21.60 &-177.74 &47.18\\ \hline
15  &1998HK151  &39.692  &0.234  &5.933  &11.880    &181.243   &50.212  &21.26 &-178.89 &44.91\\ \hline
16  &1998HQ151  &39.754  &0.290  &11.923 &20.337    &346.764   &228.831 &21.80 &-180.07      &33.76\\ \hline
17  &1998UR43   &39.302  &0.217  &8.779  &348.749   &19.006    &53.888  &21.75 &179.62 &41.40\\ \hline
18  &1998US43   &39.112  &0.131  &10.628 &48.090    &139.419   &223.893 &19.34 &178.94 &91.83\\ \hline
19  &1998VG44   &39.083  &0.249  &3.038  &350.007   &324.562   &127.946 &18.47 &179.7 &92.12\\ \hline
20  &1998WS31   &39.202  &0.196  &6.748  &8.515     &28.156    &16.008  &22.05 &179.13 &24.16\\ \hline
21  &1998WU31   &39.077  &0.184  &6.593  &34.092    &140.900   &237.186 &18.72 &177.68 &93.86\\ \hline
22  &1998WV31   &39.133  &0.271  &5.736  &53.846    &273.132   &58.527  &19.77 &178.78 &70.23\\ \hline
23  &1998WW24   &39.275  &0.223  &13.961 &30.870    &145.696   &234.005 &22.00 &175.4 &39.07\\ \hline
24  &1998WZ31   &39.346  &0.165  &14.631 &22.403    &351.955   &50.607  &21.48 &174.33 &76.61\\ \hline
25  &1999CE119  &39.583  &0.274  &1.473  &352.711   &34.967    &171.553 &18.87 &-178.94 &82.90\\ \hline
26  &1999CM158  &39.616  &0.281  &9.286  &21.325    &165.232   &338.982 &17.00 &181.32 &111.68\\ \hline
27  &1999RK215  &39.316  &0.142  &11.459 &134.601   &95.147    &137.485 &21.35 &184.25 &50.79\\ \hline
28  &1999TC36   &39.315  &0.222  &8.416  &348.380   &294.760   &97.032  &20.25 &177.65 &69.13\\ \hline
29  &1999TR11   &39.244  &0.242  &17.166 &18.660    &346.743   &54.743  &22.83 &176.6 &40.04\\ \hline
30  &2000CK105  &39.409  &0.233  &8.142  &179.861   &351.872   &326.524 &22.26 &178.76 &13.59\\ \hline
31  &2000EB173  &39.753  &0.282  &15.466 &348.858   &67.699    &169.305 &21.73 &-178.15 &22.65\\ \hline
32  &2000FB8    &39.416  &0.293  &4.580  &92.595    &67.714    &1.737	&19.55 &179.51 &73.61\\ \hline
33  &2000FV53   &39.459  &0.168  &17.306 &15.415    &351.463   &207.531 &18.94 &-176.32 &117.76\\ \hline
34  &2000GE147  &39.708  &0.237  &4.989  &5.449     &49.538    &154.709 &21.46 &-178.55 &35.76\\ \hline
35  &2000GN171  &39.694  &0.287  &10.801 &355.988   &195.189   &26.096  &21.43 &-178.85 &38.13\\ \hline
36  &2000YH2    &39.095  &0.299  &12.930 &349.766   &232.895   &219.465 &18.97 &181.47 &85.67\\ \hline
37  &2001FL194  &39.531  &0.178  &13.687 &14.089    &171.983   &2.081	&21.10 &-180.27 &85.56\\ \hline
38  &2001FR185  &39.482  &0.192  &5.634  &326.437   &334.190   &287.623 &17.61 &-178.91 &107.64\\ \hline
39  &2001FU172  &39.636  &0.272  &24.694 &30.943    &135.196   &32.448  &23.17 &-186.70 &31.70\\ \hline
40  &2001KB77   &39.939  &0.290  &17.487 &335.644   &52.484    &222.994 &17.83 &-177.21 &102.82\\ \hline
41  &2001KD77   &39.820  &0.120  &2.252  &23.670    &90.536    &139.129 &19.42 &-177.19 &95.19\\ \hline
42  &2001KN77   &39.410  &0.242  &2.357  &305.744   &279.060   &45.350  &15.25 &-181.28 &120.44\\ \hline
43  &2001KQ77   &39.779  &0.159  &15.581 &314.840   &62.819    &248.476 &22.00 &-175.58 &64.38\\ \hline
44  &2001KX76   &39.691  &0.242  &19.582 &269.043   &298.714   &71.028  &21.80 &-185.50 &47.89\\ \hline
45  &2001KY76   &39.580  &0.236  &3.963  &295.335   &261.555   &90.086  &20.59 &-181.39 &58.22\\ \hline
46  &2001QF298  &39.347  &0.112  &22.368 &140.065   &42.505    &164.186 &26.55 &179.15 &31.67\\ \hline
47  &2001QG298  &39.298  &0.192  &6.494  &354.961   &208.744   &162.546 &18.67 &178.08 &95.10\\ \hline
48  &2001QH298  &39.343  &0.110  &6.712  &53.013    &168.482   &129.440 &20.95 &181.95 &64.34\\ \hline
49  &2001RU143  &39.355  &0.152  &6.528  &140.302   &18.899    &209.183 &21.13 &181.89 &58.11\\ \hline
50  &2001RX143  &39.275  &0.298  &19.282 &87.094    &239.712   &20.558  &20.49 &180.89 &61.50\\ \hline
51  &2001UO18   &39.485  &0.284  &3.672  &329.375   &47.777    &36.385  &17.47 &179.67 &99.99\\ \hline
52  &2001VN71   &39.287  &0.243  &18.692 &359.265   &1.438     &70.441  &22.76 &179.14 &49.91\\ \hline
53  &2001YJ140  &39.282  &0.290  &5.980  &358.246   &129.452   &319.434 &19.73 &179.34 &74.73\\ \hline
54  &2002CE251  &39.543  &0.272  &9.294  &347.383   &215.444   &342.554 &14.46 &-177.96 &103.83\\ \hline
55  &2002CW224  &39.185  &0.243  &5.668  &293.038   &156.088   &1.759	&21.50 &180.41 &43.34\\ \hline
56  &2002GE32   &39.569  &0.232  &15.670 &287.290   &103.320   &203.739 &20.16 &-180.41 &73.14\\ \hline
57  &2002GF32   &39.497  &0.172  &2.779  &111.703   &54.825    &44.317  &18.89 &-181.23 &91.71\\ \hline
58  &2002GL32   &39.723  &0.131  &7.070  &4.155     &192.923   &11.080  &21.80 &-179.71 &53.86\\ \hline
59  &2002GV32   &39.797  &0.198  &5.373  &349.937   &173.000   &79.151  &21.32 &-178.84 &43.97\\ \hline
60  &2002GW31   &39.413  &0.239  &2.640  &87.855    &198.315   &227.267 &19.50 &179.04 &80.27\\ \hline
61  &2002GY32   &39.716  &0.095  &1.799  &16.554    &337.273   &225.560 &22.49 &-179.65 &24.46\\ \hline
62  &2002VD138  &39.403  &0.151  &2.784  &41.170    &36.198    &315.079 &20.18 &178.95 &76.03\\ \hline
63  &2002VE95   &39.132  &0.285  &16.346 &8.236     &206.775   &199.855 &21.38 &181.71 &56.82\\ \hline
64  &2002VR128  &39.313  &0.265  &14.035 &60.630    &287.630   &23.108  &22.00 &177.26 &18.06\\ \hline
65  &2002VU130  &39.022  &0.211  &1.373  &258.236   &281.602   &267.864 &16.27 &181.94 &115.85\\ \hline
66  &2002VX130  &39.325  &0.220  &1.322  &359.559   &106.036   &296.778 &21.18 &178.52 &46.47\\ \hline
67  &2002XV93   &39.204  &0.127  &13.286 &267.105   &165.694   &19.121  &21.96 &176.46 &42.70\\ \hline
68  &2003AZ84   &39.413  &0.181  &13.596 &215.203   &15.040    &252.143 &22.84 &179.64 &44.63\\ \hline
69  &2003FB128  &39.821  &0.260  &8.867  &38.023    &306.210   &209.482 &18.96 &-179.28 &88.10\\ \hline
70  &2003FF128  &39.831  &0.221  &1.911  &333.787   &169.763   &91.731  &20.25 &-179.16 &65.21\\ \hline
71  &2003FL127  &39.337  &0.233  &3.500  &146.187   &55.981    &314.317 &20.42 &178.97 &63.50\\ \hline
72  &2003HA57   &39.648  &0.176  &27.584 &0.276     &5.641     &199.708 &25.80 &-186.74 &48.17\\ \hline
73  &2003HD57   &39.697  &0.183  &5.612  &22.980    &137.734   &34.397  &21.15 &-179.59 &53.14\\ \hline
74  &2003HF57   &39.619  &0.198  &1.422  &23.776    &123.869   &48.151  &21.10 &-178.70 &50.19\\ \hline
75  &2003QB91   &39.210  &0.197  &6.493  &132.226   &80.949    &136.756 &19.18 &182.13 &85.39\\ \hline
76  &2003QH91   &39.540  &0.152  &3.652  &108.349   &267.910   &286.690 &18.07 &182.40 &105.60\\ \hline
77  &2003QX111  &39.403  &0.146  &9.534  &85.767    &101.157   &157.498 &21.42 &183.43 &44.50\\ \hline
78  &2003SO317  &39.318  &0.276  &6.573  &42.879    &111.721   &187.304 &17.98 &179.61 &93.04\\ \hline
79  &2003SR317  &39.426  &0.168  &8.357  &50.834    &117.872   &175.647 &19.69 &181.16 &79.06\\ \hline
80  &2003TH58   &39.240  &0.088  &27.994 &15.262    &166.733   &251.400 &22.17 &175.75 &65.29\\ \hline
81  &2003UT292  &39.095  &0.292  &17.570 &332.561   &256.085   &211.037 &19.57 &181.5 &80.69\\ \hline
82  &2003UV292  &39.197  &0.214  &10.998 &46.554    &121.092   &235.689 &21.33 &177.93 &46.02\\ \hline
83  &2003VS2    &39.266  &0.072  &14.800 &4.586     &112.370   &302.667 &25.86 &179.66 &27.96\\ \hline
84  &2003WA191  &39.157  &0.232  &4.521  &13.619    &226.152   &179.282 &21.03 &179.04 &51.52\\ \hline
85  &2003WU172  &39.058  &0.254  &4.148  &338.412   &101.101   &10.422  &17.72 &179.89 &104.89\\ \hline
86  &2004DW     &39.300  &0.222  &20.594 &162.480   &72.412    &268.724 &21.07 &182.84 &67.54\\ \hline
87  &2004EH96   &39.639  &0.283  &3.128  &11.376    &301.717   &226.249 &18.35 &-178.92 &89.89\\ \hline
88  &2004EJ96   &39.795  &0.244  &9.327  &323.832   &231.714   &18.733  &18.53 &-178.74 &93.32\\ \hline
89  &2004EW95   &39.670  &0.320  &29.241 &344.317   &204.373   &25.751  &22.40 &-176.50 &53.31\\ \hline
90  &2004FU148  &39.867  &0.235  &16.623 &318.370   &81.079    &197.868 &18.82 &-178.18 &94.67\\ \hline
91  &2004FW164  &39.752  &0.163  &9.099  &359.209   &7.466     &197.971 &20.74 &-178.92      &71.70\\ \hline
92  &2005EZ296  &39.647  &0.155  &1.773  &322.078   &211.628   &24.433  &18.12 &-177.57 &103.09\\ \hline
93  &2005EZ300  &39.721  &0.243  &10.319 &311.819   &252.031   &357.124 &16.54 &-179.18 &114.11\\ \hline
94  &2005GA187  &39.653  &0.221  &18.714 &290.525   &281.974   &27.716  &21.60 &-182.86 &46.21\\ \hline
95  &2005GB187  &39.743  &0.242  &14.659 &6.049     &349.134   &217.084 &22.46 &-180.87 &35.64\\ \hline
96  &2005GE187  &39.660  &0.329  &18.222 &324.766   &85.071    &205.378 &20.93 &-179.44 &38.86\\ \hline
97  &2005GF187  &39.803  &0.262  &3.906  &337.514   &134.526   &128.952 &21.17 &-179.76 &38.06\\ \hline
98  &2005TV189  &39.286  &0.186  &34.455 &359.741   &186.066   &246.224 &28.54 &164.83 &33.49\\ \hline

\end{longtable}
}

\bibliographystyle{aa}
\addcontentsline{toc}{chapter}{Bibliography}
\bibliography{alvaro_reviewed_ac_np_v1}

\end{document}